\documentclass[showpacs,nofootinbib,groupedaddress,superscriptaddress,preprint,showkeys]{revtex4}

\usepackage[dvips]{graphicx}
\usepackage{color} 
\usepackage{epsfig}
\usepackage{amsmath}
\usepackage{amssymb}

\allowdisplaybreaks[4]

\begin{document}

\title{Signature of the Minimal Supersymmetric Standard Model with Right-Handed Neutrinos
 in Long Baseline Experiments}

\author{Toshihiko Ota}
\email{toshi@het.phys.sci.osaka-u.ac.jp}
\affiliation{Department of Physics, Osaka University,
        Machikaneyama, Toyonaka, Osaka, 560-0043, Japan}

\author{Joe Sato}
\email{joe@phy.saitama-u.ac.jp}
\affiliation{Department of Physics, Saitama University, 
        Shimo-okubo, Sakura-ku, Saitama, 338-8570, Japan}
\affiliation{Physik-Department Technische Universit\"at M\"unchen,
       James-Franck-Strasse 85748 Garching, Germany}

\preprint{OU-HET-517}
\preprint{STUPP-05-179}
\preprint{TUM-HEP-574/05}



\begin{abstract}
The effective interactions which violate a lepton flavor accompanied 
with neutrinos (nLFV) are considered.
Such a new physics effect is expected to be measured 
in future neutrino oscillation experiments with long baseline.
They are induced by radiative correction in the framework of 
the minimal supersymmetric standard model with right-handed neutrinos.
We numerically evaluate the size of the couplings  for nLFV interactions 
in this framework.
The slepton mixing is not only the origin of the lepton flavor
violation in the charged lepton sector (cLFV) but also that of the
nLFV.
We find that the nLFV couplings are strongly correlated with the corresponding 
cLFV process, and they are constrained at
$\mathcal{O}(10^{-5})$ times smaller than the standard four-Fermi
couplings.
\end{abstract}

\maketitle

\section{Introduction}
Numerous observations on neutrinos from 
the sun\cite{solar-r1,solar-r2,solar-r3,solar-r4,solar-sk,solar-sno}, 
the atmosphere\cite{atm1,atm2,atm3}, 
the reactor\cite{reactor},
and the accelerator\cite{K2K} suggest that
neutrinos are massive and hence there are mixings in the lepton sector.
This fact means that the standard model (SM) has to be extended
so that the neutrino masses and the lepton mixings are introduced 
into the model.
Lots of models to explain those experimental results have been proposed. 
Among them, a model with the seesaw mechanism\cite{seesaw} 
has a promising attribute, in which tiny neutrino masses are naturally 
induced. 
%
Neutrino experiments also have revealed that the mixings in the lepton
sector are much larger than those in the quark sector. This fact may imply
that the lepton flavor number is strongly violated in the physics beyond
the SM.  Therefore, we can expect that the nature might exhibit sizable
lepton flavor violation (LFV) and hence we could observe the remains of
physics at the high energy scale. In the minimal supersymmetric standard model with
heavy right-handed neutrinos (MSSMRN), in which the seesaw mechanism is
realized, the LFV with charged leptons (cLFV) is expected to become
large\cite{BM,HMTYY,MSSMRNcLFV}. In this class of models, the renormalization
effect due to
the neutrino Yukawa couplings
induces a significant size of off-diagonal elements of the slepton mass matrix,
$(m^2_{\tilde L})_\alpha^{\ \beta}$ ($\alpha\ne\beta$, $\alpha,\beta = e, \mu, \tau$),
which are the seeds of the cLFV.
Here, the flavor indices $\alpha$ and $\beta$ should be understood 
to also indicate the mass eigenstates of the charged lepton fields.
Concretely, 
the superpotential with the neutrino Yukawa couplings ${(f_{\nu})_{i}}^{\alpha}$
and the Majorana masses for the right-handed neutrinos $M^{ij}$
includes the following terms,
\begin{eqnarray}
W \supset  {(f_{\nu})_{i}}^{\alpha} \bar{N}^{i} H_{u} \epsilon L_\alpha  
 + 
 \frac{1}{2} M^{ij}N_iN_j,
\label{eq:DiracYukawa}
\end{eqnarray}
where $N_i$, $L_\alpha$, and $H_{u}$ denote the chiral supermultiplet for 
the right-handed neutrinos, 
that for the lepton doublets, and 
that for the up-type Higgs field, 
and $\epsilon$ is the anti-symmetric tensor for $SU(2)_{L}$ fundamental 
representation.
The indices $i$ and $j$ are for the generation of right-handed neutrinos, 
which do not necessarily indicate the mass eigenstates.
The renormalization group equation for the off-diagonal elements of the
slepton mass matrix is given as\cite{BM},
\begin{eqnarray}
\mu \frac{{\rm d}(m^2_{\tilde{L}})_\alpha^{\ \beta}}{{\rm d}\mu}
=
\frac{1}{16\pi^2} \left[
m^2_{\tilde{L}} f_\nu^\dagger f_\nu+f_\nu^\dagger f_\nu m^2_{\tilde{L}}
+2(f_\nu^\dagger m^2_{\tilde \nu} f_\nu +\tilde{m}^2_{H_u}
f_\nu^\dagger f_\nu +A_\nu^\dagger A_\nu)\right]_\alpha^{\ \beta}, 
\quad (\alpha \neq \beta),
\label{eq:RGEforL}
\end{eqnarray}
where 
$m^2_{\tilde{\nu}}$ is the soft supersymmetry (SUSY) breaking mass matrix 
for the right-handed sneutrino, 
and
$\tilde{m}^2_{H_u}$ is that for the up-type Higgs doublet.
The matrix $A_\nu$ denotes the tri-linear scalar couplings corresponding to
the first term in Eq. \eqref{eq:DiracYukawa}.  
Note that if the neutrino Yukawa couplings 
do not exist, then there is no LFV effect.
We can approximately solve Eq. \eqref{eq:RGEforL} as
\begin{eqnarray}
{(\Delta m^2_{\tilde{L}})}_\alpha^{\ \beta}
\simeq -\frac{(6+ 2 a_0^2) m_0^2}{16 \pi^2}
(f_\nu^\dagger f_\nu)_\alpha^{\ \beta} \ln \frac{M_G}{M_R},
\label{eq:RGESolution}
\end{eqnarray}
with a cutoff scale $M_G$ and a typical mass scale for 
right-handed neutrinos $M_R$. 
Here, the universal soft SUSY breaking at $M_{G}$ is assumed, 
and $m_{0}$ is the parameter for sfermion masses and $a_{0}$ is that
for scalar tri-linear couplings.
\begin{figure}[tbh]
\unitlength=0.9cm
\begin{picture}(10,7)
\thicklines
\put(0.5,4){\line(1,0){2}}
\put(1.5,4){\vector(1,0){0}}
\put(0.6,4.2){$l_{\beta}$ }
\put(2.5,4){\circle*{0.2}}
\multiput(2.5,4)(0.5,0){8}{\line(1,0){0.3}}
\put(3.5,4){\vector(1,0){0}}
\put(3.2,4.2){$\tilde{l}_{\beta}$}
%
\put(4,3.2){${(\Delta m^2_{\tilde{L}})_{\alpha}}^{\beta}$}
\put(4.5,4){\circle*{0.2}}
\put(5.5,4){\vector(1,0){0}}
\put(5.2,4.2){$\tilde{l}_{\alpha}$}
\put(6.5,4){\circle*{0.2}}
\put(6.5,4){\line(1,0){2}}
\put(8,4){\vector(1,0){0}}
\put(8,4.2){$l_{\alpha} $}
\put(4.5,4){
\qbezier(-2,0)(-2.01,0.35)(-1.88,0.68)
\qbezier(-1.88,0.68)(-1.78,1.03)(-1.53,1.29)
\qbezier(-1.53,1.29)(-1.3,1.55)(-1,1.73)
\qbezier(-1,1.73)(-0.69,1.9)(-0.35,1.97)
\qbezier(-0.35,1.97)(0,2.03)(0.35,1.97)
\qbezier(2,0)(2.01,0.35)(1.88,0.68)
\qbezier(1.88,0.68)(1.78,1.03)(1.53,1.29)
\qbezier(1.53,1.29)(1.3,1.55)(1,1.73)
\qbezier(1,1.73)(0.69,1.9)(0.35,1.97)
}
\put(4.75,6){\vector(1,0){0}}
\put(4.5,4){
\qbezier(-2,0)(-2.36,0.417)(-1.88,0.68)
\qbezier(-1.88,0.68)(-1.47,0.85)(-1.53,1.29)
\qbezier(-1.53,1.29)(-1.54,1.84)(-1,1.73)
\qbezier(-1,1.73)(-0.58,1.6)(-0.35,1.97)
\qbezier(-0.35,1.97)(0,2.35)(0.35,1.97)
\qbezier(2,0)(2.36,0.417)(1.88,0.68)
\qbezier(1.88,0.68)(1.47,0.85)(1.53,1.29)
\qbezier(1.53,1.29)(1.54,1.84)(1,1.73)
\qbezier(1,1.73)(0.58,1.6)(0.35,1.97)
}
\put(4.4,6.5){$\tilde{\chi}^{0}$}
\put(2.5,-1.2){
\rotatebox{50}{
\multiput(5.2,0.5)(0,-1){3}
{\qbezier(0,0)(-0.25,-0.25)(0,-0.5)
 \qbezier(0,-0.5)(0.25,-0.75)(0,-1.0)}
}
}
\put(6.2,1.8){$\gamma$}
\end{picture}\vspace{-1.4cm}
\caption{One of the diagrams which contribute to the cLFV process 
$l_\beta \rightarrow l_{\alpha} \gamma$.
This effect is approximately understood by the insertion of the LFV mass term 
$(\Delta m^2_{\tilde{L}})_\alpha^{\ \beta}$ $(\alpha\ne\beta)$.}
\label{fig:ChargedLFV}
\end{figure}
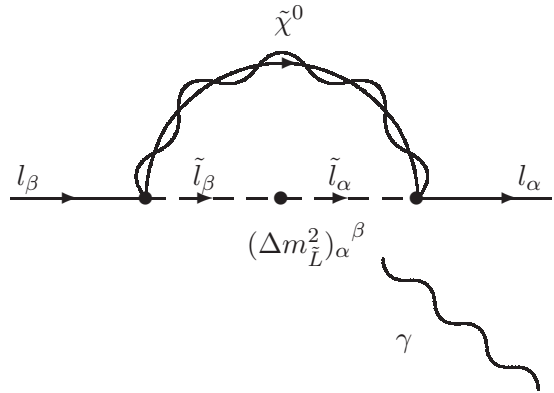
In terms of the mass insertion method,
we can see that the off-diagonal elements of the slepton mass matrix
are the origins of cLFV.  
This is diagrammatically shown in Fig. \ref{fig:ChargedLFV}.
From this diagram, it is obvious that 
the element {$(\Delta m^2_{\tilde{L}})_\alpha^{\ \beta}$} is relevant
to the process
{$l_{\beta} \rightarrow l_{\alpha} \gamma$}.
In this class of models, 
the off-diagonal elements can become
large so that the typical values of predicted branching ratios are 
within a sensitivity reach of near future experiments\cite{PSI,MECO,PRISM}.
Therefore, the search for the cLFV process is one of the promising
ways to inspect the new physics effect beyond the SM.  

In this article, however, we investigate an alternative approach to explore
the LFV, the search for the processes of the LFV with neutrinos (nLFV)
at a long baseline (LBL) neutrino oscillation experiment. In the
forthcoming experiments, the oscillation parameters such as the mixing
angles and the squared mass differences are expected to be determined
with high precision\cite{T2K,nuFact,combined}. Therefore, the measurement of 
nLFV effects might be possible. 
The feasibility studies on the nLFV interaction search at future LBL
experiments without assuming a specific model are made by
Refs. \cite{Grossman,GGGN,NewPhysMatter,HV,HSV,OSY,OS,Campanelli,Tokushima,Melbourne}.
The current experimental bounds on nLFV interactions are given in
Ref. \cite{Davidson}.
The sensitivity for nLFV effects to  
solar neutrinos\cite{NSIinSolar}, atmospheric
neutrinos\cite{NSIinAtmos}, the LSND results\cite{NSIinLSND} 
and supernova neutrinos\cite{NSIinSN} has also been considered. 
It is pointed out that the nLFV signal is enhanced by the interference
effect between the amplitude including the nLFV interactions and that of the
standard oscillation (SO).

We here investigate the nLFV interactions in the MSSMRN\footnote{%
The studies for the other models are done in e.g., Ref. \cite{Branco}.}.
The origins of nLFV processes are the same as those of cLFV
processes, which are the off-diagonal elements of the slepton mass
matrix.
They can become significant in this framework.
In addition, there is an enhancement mechanism due to the interference effect. 
It can be expected that the detectable magnitude of nLFV effects is induced.
We make the numerical calculations of the size of the nLFV couplings and show
the correlation between the nLFV and the cLFV.

In Sec. \ref{sec:paramtrization}, we recapitulate the model independent
approach in the detection of nLFV effects at LBL neutrino
oscillation experiments.
We also show the way to parameterize the nLFV interactions. 
In Sec. \ref{sec:MSSM}, we
calculate these nLFV couplings in the MSSMRN
and numerically evaluate the size of them under the universal soft SUSY
breaking scenario, so-called the constrained MSSMRN.  
Here, we concentrate our attention on the nLFV interactions 
which are relevant in the oscillation channel $\nu_{\mu} \rightarrow \nu_{\tau}$.
Finally, in Sec. \ref{sec:summary} we will give a summary.
In Appendix \ref{app:Model},
we describe the model in order to make our notation clear,
and in Appendix \ref{app:calculation} we give formulae
of the nLFV interactions.

\section{nLFV interaction in neutrino oscillation}
\label{sec:paramtrization}

In this section, we explain how to parameterize the new physics
effect with the model independent way\cite{OSY}.
First, we note that in neutrino oscillation experiments 
we do not observe neutrinos themselves but do their products, 
corresponding charged leptons.
Propagating neutrinos appear only in intermediate states. 
Therefore, the existence of the nLFV interactions suggests that
there are some amplitudes whose initial and final states are the same as
those of the SO which means the neutrino oscillation with SM interactions.

To make the argument clear, we show an example.  
When we assume $\nu_{\mu} \rightarrow \nu_{\tau}$ oscillation
measurement at a neutrino factory experiment, 
all we can know are the facts, the decay of muons at
a muon storage ring and the appearance of tau leptons in a detector 
which is located at a length $L$ away from the source of the neutrino
beam just after the time $L/c$, where $c$ is speed of light.  
We interpret these events as
the evidence of the SO, $\nu_\mu\rightarrow\nu_\tau$. The amplitude for
this process $A_{\rm SO}$ can be expressed by the product of the
amplitudes for the sub-processes;
\begin{eqnarray}
A_{\rm SO}(\mu^{-} + I \rightarrow \tau^{-} + F)  
=
A_s(\mu^{-} \rightarrow \nu_{\mu} \bar{\nu}_{e} e^{-}) 
A_p(\nu_{\mu} \xrightarrow[]{\text{osc.}} \nu_{\tau})
A_d(\nu_{\tau}  d \rightarrow \tau^{-} u), 
\label{eq:SMAmplitude}
\end{eqnarray}
where $I$ ($F$) denotes all the other particles than a muon (tau
lepton) in the initial (final) state, which can be measured in
principle.
In this example, $I$ is $d$ (a down-type quark) in $A_d$, and 
$F$ means $\bar{\nu}_{e}$\footnote{%
Precisely, it is one of the mass eigenstate of the
neutrino\cite{Giunti}.%
} and $e^{-}$ in $A_s$ and $u$ (a up-type
quark) in $A_d$. The subscripts, $s, p$, and $d$, indicate ``at the {\it
source} of the neutrino beam'', ``at the {\it propagation} process'', and
``at the {\it detection} process'', respectively.
Suppose that there is an effective four-Fermi nLFV interaction, 
\begin{equation}
\mathcal{L}_{\text{eff}} =
\lambda (\bar{e} \gamma^{\rho} {\rm P}_{L} \mu) 
(\bar{\nu}_{\tau} \gamma_{\rho} {\rm P}_{L} \nu_{e})
+\ \hbox{\rm h.c.},
\label{eq:newPhysicsAtCreation}
\end{equation}
we have the same signal through the other process than Eq. \eqref{eq:SMAmplitude};
\begin{eqnarray}
A_{\rm nLFV}(\mu^{-} + I \rightarrow \tau^{-} + F)  
=
A_{s}(\mu^{-} \rightarrow \nu_{\tau} \bar{\nu}_{e} e^{-}) 
A_{p}(\nu_{\tau} \xrightarrow[]{\text{no osc.}} \nu_{\tau})
A_{d}(\nu_{\tau}  d \rightarrow \tau^{-} u). 
\label{eq:nLFVAmplitude}
\end{eqnarray}
The external particles in Eq. \eqref{eq:nLFVAmplitude} are completely the
same as those in Eq. (\ref{eq:SMAmplitude}). 
Therefore, we can not distinguish the contributions from these two
amplitudes. In quantum mechanics,
we first sum up these amplitudes and next
square the summation in order to obtain the transition rate. 
Therefore, an interference term arises between these two amplitudes 
for this process\footnote{%
It is necessary to treat the neutrino as a wave packet\cite{wavepacket} 
in the discussion on the coherence between these two amplitude.
Here, we adopt {\it usual} treatment for the neutrino propagation, 
so that the neutrino propagation is described by the plane wave.
}:
\begin{eqnarray}
P({\mu^{-} + I \rightarrow \tau^{-} + F})
= |A_{\rm SO}|^{2} +
2 {\rm Re}[A_{\rm SO}^{*} A_{\rm nLFV}] + |A_{\rm nLFV}|^{2} .
\label{eq:Interference}
\end{eqnarray}
The term of the SO, the first term of the right-hand side, gives the
leading contribution, and it is described by using the muon decay width
$\Gamma$ $(= |A_{s}(\mu^{-} \rightarrow \nu_{\mu} \bar{\nu}_{e} e^{-})|^{2})$ 
and the cross-section for the charged current interaction
$\sigma$ $(= |A_{d} (\nu_{\tau} d \rightarrow \tau^{-} u) |^{2})$ as
\begin{align}
\left|A_{\rm SO} \right|^{2} = \Gamma \times P_{\nu_{\mu} \rightarrow
 \nu_{\tau}} \times \sigma,
\end{align}
where $P_{\nu_{\mu} \rightarrow \nu_{\tau}}$ is the oscillation
probability for $\nu_{\mu} \rightarrow \nu_{\tau}$ in the SO, 
which is defined as $|A_{p}(\nu_{\mu} \xrightarrow[]{\text{osc.}} \nu_{\tau})|^{2}$.
The second term which represents the interference between the amplitude
of the SO Eq. (\ref{eq:SMAmplitude}) and that including the
nLFV interaction Eq. (\ref{eq:nLFVAmplitude}):
\begin{align}
2\text{Re}[A_{\rm SO}^{*} A_{\text{nLFV}}]
=
\Gamma
\times
2 \text{Re}\left[ 
\frac{\lambda}{2 \sqrt{2} G_{F}} 
A_{p}^{*}(\nu_{\mu}\xrightarrow[]{\text{osc.}} \nu_{\tau})
A_{p}(\nu_{\tau} \xrightarrow[]{\text{no osc.}} \nu_{\tau})
\right]
\times \sigma,
\end{align}
where $G_F$ is the Fermi constant.
Note that the nLFV effect contributes to the oscillation probability
not quadratically but linearly. Thus the effect can be {\it enhanced}
and hence even if the nLFV coupling is small, 
it can contribute the oscillation probability significantly\cite{HSV,OSY,OS}.

We now turn to the parametrization of effective couplings for nLFV interactions.
As we have already shown, the amplitude for {\it the neutrino
oscillation process} 
can be divided into three pieces, $A_s$, $A_p$, and $ A_d$.
First, we consider the decay process of parent particles, $A_s$.  
Since all final states must be the same, 
nLFV interactions with $(V-A)(V-A)$ type  are important for the neutrino
factory experiment\cite{OSY}.  
We can introduce the interference effect by treating the
initial state of a propagating neutrino as the superposition of all flavor
eigenstates.  For the case of Eq. \eqref{eq:newPhysicsAtCreation}, 
we can take the initial neutrino state $|\nu \rangle$ as
\begin{eqnarray}
|{\nu} \rangle = |{\nu}_{\mu} \rangle 
 + \epsilon^s_{\mu\tau} |{\nu}_{\tau} \rangle .
\end{eqnarray}
where $\epsilon^s_{\mu\tau} = \lambda/ (2\sqrt{2} G_{F})$.
It can be generalized to the case of an initial neutrino with
any flavor by using the source state notation which is introduced 
in Ref. \cite{Grossman} as\footnote{%
The matrix $U^s$ is not necessarily unitary.}
\begin{eqnarray}
 |\nu^{s}_{\beta} \rangle = {(U^{s})_{\beta}}^{\alpha} |\nu_{\alpha}\rangle , 
\qquad 
 U^{s} \equiv \left(
  \begin{array}{ccc}
   1 & \epsilon^{s}_{e \mu} & \epsilon^{s}_{e \tau}    \\
   \epsilon^{s}_{\mu e} & 1 &\epsilon^{s}_{\mu \tau}   \\
   \epsilon^{s}_{\tau e} & \epsilon^{s}_{\tau \mu} & 1
  \end{array}\right) . 
\label{eq:def-source-state}
\end{eqnarray}
We can include the total nLFV effect into the oscillation
probability as
\begin{eqnarray}
 P_{\nu^{s}_{\alpha} \rightarrow \nu_{\beta}} &=
  \left|
   \langle \nu^{\beta} |
   {\left( {\rm e}^{-{\rm i} H_{\rm SO} L} \right)_{\beta}}^{\alpha}
    {\left(U^{s}\right)_{\alpha}}^{\gamma} | \nu_{\gamma} \rangle  
  \right|^{2},
\label{eq:probability-for-V-A}
\end{eqnarray}
with the propagation Hamiltonian for the SO $H_{\rm SO}$;
\begin{align}
{(H_{\rm SO})_{\beta}}^{\alpha} 
= \frac{1}{2 E_{\nu}}
     \left\{
       {\left(U_{\rm MNS}\right)_{\beta}}^{i}\left(
        \begin{array}{ccc}
            0 & & \\
              & \Delta m_{21}^{2} & \\
              &                   & \Delta m_{31}^{2}
        \end{array}\right)
       {\left( U_{\rm MNS}^{\dagger} \right)_{i}}^{\alpha}
              +    
              {\left(
       \begin{array}{ccc}
         \bar{a}  
                &    
                &   \\
                & 0  
                &   \\
                &   
                & 0 
        \end{array}\right)_{\beta}}^{\alpha}
       \right\},
\label{eq:Hamiltonian}
\end{align}
where $E_{\nu}$ is the neutrino energy,
$\bar{a}$ is the usual matter effect which is given as $2\sqrt{2} G_{F} n_{e}
E_{\nu}$ by using the electron number density $n_{e}$, 
$U_{\rm MNS}$ is the mixing matrix for the lepton sector, 
and $\Delta m^{2}_{21}$ ($\Delta m^{2}_{31}$) is the mass squared
difference for the solar (atmospheric) neutrino oscillation.

Next, we consider the propagation process, $A_p$.
The nLFV interactions modify the Hamiltonian for neutrino propagation
as\cite{NewPhysMatter},
\begin{eqnarray}
{(H_{\rm nLFV})_{\beta}}^{\alpha} 
= 
{(H_{\rm SO})_{\beta}}^{\alpha} 
+
\frac{1}{2 E_{\nu}}
{\begin{pmatrix}
      a_{e e} 
                & a_{e \mu}   
                & a_{e \tau}  \\
              a_{e \mu}^{*}  
                & a_{\mu \mu}  
                & a_{\mu \tau}  \\
              a_{e \tau}^{*} 
                & a_{\mu \tau}^{*}  
                & a_{\tau \tau} 
 \end{pmatrix}_{\beta}}^{\alpha}
,
\end{eqnarray}
where 
$a_{\alpha \beta}$ is the extra matter effect due to 
nLFV interactions, which is defined as 
$
 a_{\alpha \beta} = 
\sum_{p=e,d,u}2\sqrt{2}
\epsilon^{m,p}_{\alpha \beta} G_{F} n_{p} E_{\nu}, 
$
where $n_p$ is the number density for the particle $p$.
Assuming the matter which consists of the same number of 
electron, neutron and proton, we can reduce it to
\begin{gather}
a_{\alpha \beta} = 2\sqrt{2} \epsilon^{m}_{\alpha\beta} G_{F}
 n_{e} E_{\nu},
\end{gather}
with
\begin{gather}
\epsilon^{m}_{\alpha \beta} \equiv
 \epsilon^{m,e}_{\alpha \beta}
 +
 3 \epsilon^{m,d}_{\alpha \beta}
 +
 3 \epsilon^{m,u}_{\alpha \beta}.
\end{gather}
Note that to
consider the magnitude of the matter effect, 
the type of the interaction is irrelevant 
since matter particles are at rest and hence the
dependence on their chirality is averaged out\cite{BGE}.

Then, we make a comment on nLFV interactions which affect a detection
process, $A_d$. 
We can adopt a quite similar treatment to that at the source of the neutrino beam.
In this article, we consider the case in which the nLFV interactions do not
depend on the flavor of target-quark, which is almost the case for the
so-called constrained MSSMRN. 
Therefore, we have the neutrino state for the detection process in the
following form,
\begin{eqnarray}
 |\nu^{d}_{\beta} \rangle = {(U^{d})_{\beta}}^{\alpha} |\nu_{\alpha}\rangle , 
  \qquad
 U^{d} \equiv \left(
  \begin{array}{ccc}
   1 & \epsilon^{d}_{e \mu} & \epsilon^{d}_{e \tau}    \\
   \epsilon^{d}_{\mu e} & 1 &\epsilon^{d}_{\mu \tau}   \\
   \epsilon^{d}_{\tau e} & \epsilon^{d}_{\tau \mu} & 1
  \end{array}\right) . 
\label{eq:def-detection-state}
\end{eqnarray}

Finally, the transition probability including the whole nLFV effects
is given as 
\begin{align}
P(\mu^{-} +I \rightarrow \tau^{-} + F)
\simeq 
 \Gamma \times  P_{\nu^{s}_{\alpha} \rightarrow \nu_{\beta}^{d}}
 \times \sigma,
\end{align}
where
\begin{eqnarray}
 P_{\nu^{s}_{\alpha} \rightarrow \nu_{\beta}^{d}} & =
  \left|
   \langle \nu^{\delta} | {(U^{d\dagger})_{\delta}}^{\beta} 
   {({\rm e}^{-{\rm i} H_{\text{nLFV}} L})_{\beta}}^{\alpha}
   {(U^{s})_{\alpha}}^{\gamma} | \nu_{\gamma} \rangle  
  \right|^{2}.
\label{eq:probability}
\end{eqnarray}

\section{nLFV interactions in the MSSMRN}
\label{sec:MSSM}

\begin{figure}[tbh]
\unitlength=0.9cm
\begin{picture}(10,7)
\thicklines
\put(0.5,4){\line(1,0){2}}
\put(1.5,4){\vector(1,0){0}}
\put(0.5,4.3){$l_{\beta}$}
\put(2.5,4){\circle*{0.2}}
\multiput(2.5,4)(0.5,0){8}{\line(1,0){0.3}}
\put(3.5,4){\vector(1,0){0}}
\put(3.1,4.3){$\tilde{l}_{\beta}$}
\put(4.5,4.3){$\tilde{l}_{\alpha}$}
\put(5.4,4){\circle*{0.2}}
\put(3.8,4){\circle*{0.2}}
\put(6,4){\vector(1,0){0}}
\put(4.7,4){\vector(1,0){0}}
\put(5.7,4.3){$\tilde{\nu}_{\alpha}$}
\put(6.5,4){\circle*{0.2}}
\put(6.5,4){\line(1,0){2}}
\put(8,4){\vector(1,0){0}}
\put(8,4.3){$\nu_{\alpha} $}
\put(4.5,4){
\qbezier(-2,0)(-2.01,0.35)(-1.88,0.68)
\qbezier(-1.88,0.68)(-1.78,1.03)(-1.53,1.29)
\qbezier(-1.53,1.29)(-1.3,1.55)(-1,1.73)
\qbezier(-1,1.73)(-0.69,1.9)(-0.35,1.97)
\qbezier(-0.35,1.97)(0,2.03)(0.35,1.97)
\qbezier(2,0)(2.01,0.35)(1.88,0.68)
\qbezier(1.88,0.68)(1.78,1.03)(1.53,1.29)
\qbezier(1.53,1.29)(1.3,1.55)(1,1.73)
\qbezier(1,1.73)(0.69,1.9)(0.35,1.97)
}
\put(4.8,6){\vector(1,0){0}}
\put(4.5,4){
\qbezier(-2,0)(-2.36,0.417)(-1.88,0.68)
\qbezier(-1.88,0.68)(-1.47,0.85)(-1.53,1.29)
\qbezier(-1.53,1.29)(-1.54,1.84)(-1,1.73)
\qbezier(-1,1.73)(-0.58,1.6)(-0.35,1.97)
\qbezier(-0.35,1.97)(0,2.35)(0.35,1.97)
\qbezier(2,0)(2.36,0.417)(1.88,0.68)
\qbezier(1.88,0.68)(1.47,0.85)(1.53,1.29)
\qbezier(1.53,1.29)(1.54,1.84)(1,1.73)
\qbezier(1,1.73)(0.58,1.6)(0.35,1.97)
}
\put(4.3,6.5){$\tilde{\chi}^{0}$}
\rotatebox{30}{
\multiput(6.2,1)(0,-1){3}
{\qbezier(0,0)(-0.25,-0.25)(0,-0.5)
 \qbezier(0,-0.5)(0.25,-0.75)(0,-1.0)}
}
\put(4.3,2){$W$}
\put(5.4,1.4){\line(1,0){2.5}}
\put(5.4,1.4){\line(4,-1){2.5}}
\put(5.4,1.4){\circle*{0.2}}
\put(6.9,1.4){\vector(-1,0){0}}
\put(6.92,1.02){\vector(4,-1){0}}
\put(7.6,1.6){$\bar{\nu}_{e}$}
\put(6.9,0.6){$e$}
\put(1.5,3.2){${(\Delta m^2_{\tilde{L}})_{\alpha}}^{\beta}$}
\end{picture}\vspace{-0.8cm}
\caption{One of the diagrams which contribute to $\epsilon^{s}_{\beta \alpha}$.}
\label{fig:diagram-for-epsilon}
\end{figure}
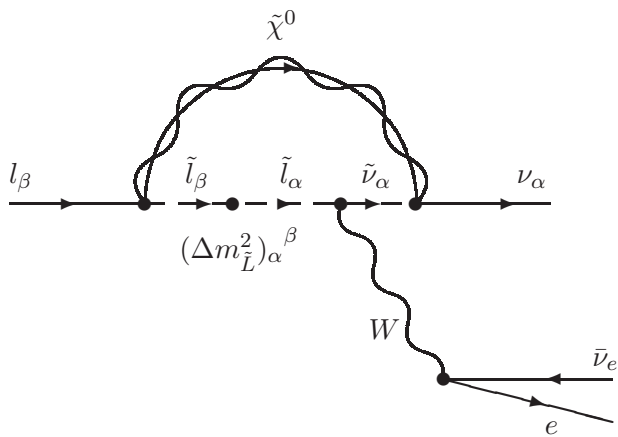

In this section, we evaluate the effective couplings for 
nLFV interactions in the MSSMRN.  The nLFV interactions are induced by
the off-diagonal elements of the slepton mass matrix, which are similar
to the cLFV interactions.  We can naively estimate the size
of the nLFV parameter $\epsilon^{s}_{\alpha\beta}$ from the diagram
which is shown in Fig. \ref{fig:diagram-for-epsilon} as\cite{osNufact04}
\begin{align}
\epsilon^{s}_{\alpha \beta}
 \left(
 \sim \epsilon^{s}_{\beta \alpha}
 \right)
 \sim
\frac{g^{2}_{2}}{16\pi^2}\frac{{(\Delta m^2_{\tilde L})_{\beta}}^{\alpha}}
{m_{\rm S}^2}
\sim 
 m_{\rm S}^2 \times \sqrt{\text{Br}(l_{\beta} \rightarrow
 l_{\alpha}\gamma)}.
\label{eq:naive-estimation-epsilon}
\end{align}
Here $g_{2}$ is the gauge coupling for $SU(2)_{L}$ and
$m_{\rm S}$ is the typical SUSY breaking scale.
This relation means that there is a correlation between two processes,
$\l^{-}_{\beta} \rightarrow \nu_{\alpha} \bar{\nu}_{e} e^{-}$ and
$l_\beta \rightarrow l_\alpha \gamma$.
We here concentrate our attention on the nLFV associated 
with the tau lepton
because that in $\mu$-$e$ sector is strongly constrained by
corresponding cLFV processes. 
The current experimental bound on the branching ratio of
$\tau\rightarrow \mu \gamma$ is $3.1 \times 10^{-7}$ at 90\% confidence
level\cite{Belle-tmg}, and then this experimental bound constrains the
nLFV coupling parameter $\epsilon^{s}_{\mu\tau}$.  
According to this naive estimation, the value of
$\epsilon^{s}_{\mu\tau}$ may become as large as $\mathcal{O}(10^{-4})$.
Such size of nLFV interactions would be detected 
at a future LBL experiment such as the neutrino
factory\cite{OSY}. 

\subsection{Analytic Calculation of $\epsilon^{s,m,d}_{\alpha\beta}$}
\label{sec:analytic-calc}

In this subsection, we explain the calculation to
obtain $\epsilon^{s,m,d}_{\alpha \beta}$ in detail 
and compare them with that of cLFV processes.
The thorough results are given in Appendix \ref{app:calculation}.

For example, the effective Lagrangian relevant with 
the nLFV interactions which give potentially significant
contribution to the oscillation $\nu_{\mu}
\rightarrow
\nu_{\tau}$ in a neutrino factory is given as\cite{OSY}
\begin{align}
\mathcal{L}^{\text{nLFV}}_{\text{eff}}
 =&
 2\sqrt{2} G_{F} \epsilon^{s}_{\mu\tau}
 \left(
 \bar{\nu}_{\tau} \gamma^{\rho} {\rm P}_{L} \nu_{e}
 \right)
 \left(
 \bar{e} \gamma_{\rho} {\rm P}_{L} \mu
 \right) \nonumber \\
 &+
 2\sqrt{2} G_{F} \sum_{p=e,d,u} \epsilon^{m,p}_{\mu\tau}
 \left(
 \bar{\nu}_{\tau} \gamma^{\rho} {\rm P}_{L} \nu_{\mu}
 \right)
 \left(
 \bar{p} \gamma_{\rho} p
 \right) \\
 &+
 2\sqrt{2} G_{F} \epsilon^{d}_{\mu\tau}
 \left(
 \bar{\tau} \gamma^{\rho} {\rm P}_{L} \nu_{\mu}
 \right)
 \left(
 \bar{u} \gamma_{\rho} {\rm P}_{L} d
 \right). \nonumber 
\end{align}
These effective couplings arise from penguin-type diagrams
and from box-type diagrams as shown in Appendix \ref{app:calculation}.
The calculation of box diagrams is straightforward.
It is almost the same as that of cLFV processes, e.g., $\mu$-$e$
conversion\cite{HMTYY} except for the fact that 
only the $(V-A)(V-A)$ type interactions are taken into account
in the calculation of $\epsilon^{s}_{\alpha \beta}$ and $\epsilon^{d}_{\alpha\beta}$.
However, it is necessary to make an attentive calculation for 
penguin-type diagrams. 
In general, while the neutral and electromagnetic currents corresponding to
Fig. \ref{fig:ChargedLFV} take the following form,
\begin{align}
 &\bar{f}^\alpha (p-q) 
  \left[
   {\rm i} 
   \left\{ 
    \mathfrak{A}(q^{2})+\mathfrak{B}(q^{2}) \gamma_5
   \right\} q_\nu\sigma^{\mu\nu}
   + 
   \left\{
    \mathfrak{C}(q^{2})+\mathfrak{D}(q^{2}) \gamma_5
   \right\} \gamma^{\mu}
   + 
   \left\{ 
    \mathfrak{E}(q^{2})+\mathfrak{F}(q^{2}) \gamma_5
   \right\} q^{\mu}
   \right] f_\beta (p) \nonumber \\ & + {\rm h.c.}, 
\end{align}
the charged current corresponding to Fig. \ref{fig:diagram-for-epsilon} 
is decomposed to 
\begin{eqnarray}
 \bar{\nu}^\alpha (p-q) {\rm P}_{R}
  \left\{ {\rm i} \mathfrak{A}'(q^{2}) q_\nu \sigma^{\mu\nu}
   + \mathfrak{C}'(q^{2}) \gamma^\mu
   + \mathfrak{E}'(q^{2}) q^\mu
   \right\}  
  l_\beta (p) + {\rm h.c.}.
\end{eqnarray}
Here, $f$ denotes the charged lepton field $l$ or the neutrino field
$\nu$, $p$ is the momentum for the incoming particle, and
$q$ is that for the gauge boson.
All the coefficients, $\mathfrak{A}$ ($\mathfrak{A}'$), $\mathfrak{B}$,
$\mathfrak{C}$ ($\mathfrak{C}'$),
$\mathfrak{D}$, $\mathfrak{E}$ ($\mathfrak{E}'$), 
and $\mathfrak{F}$, are the functions of $q^2$.
In the limit $q^2\rightarrow 0$, 
$\mathfrak{C}$ and $\mathfrak{D}$ 
for the electromagnetic current must vanish due to the gauge symmetry
$U(1)_{\rm em}$.
On the other hand, those for the neutral and the charged current 
do not vanish because the corresponding gauge
symmetry $SU(2)_{L}$ is broken.  
In other words, the Lorentz structure of the nLFV interaction 
which is shown in Fig. \ref{fig:diagram-for-epsilon} is different from
that of the cLFV process $l_{\beta} \rightarrow l_{\alpha} \gamma$ with
a real photon emission.  The former one is dominated by the vector
exchange interaction ($\mathfrak{C}'(0)$),\footnote{%
Since the effective four-Fermi
couplings are induced by the exchange of the massive gauge bosons $Z$
and $W$ and $q^2$ is of $\mathcal{O}(m_f)$, essentially we can put $q^2=0$.} 
while the latter is done by a di-pole type interaction 
($\mathfrak{A}(0)+\mathfrak{B}(0)\gamma_5$).
Therefore, they are definitely correlated each other but are not the same
functions.
\begin{figure}[htb]
\unitlength=0.8cm
\begin{picture}(10,7)
\thicklines
\put(1.8,4){\color{red}\oval(2,0.5)}
\put(-0.5,4){\line(1,0){3}}
\put(0.5,4){\vector(1,0){0}}
\put(-0.7,4.3){$l_{\beta}$}
\put(-0.7,3.5){$p \hspace{0.1cm}$}
\put(-0.3,3.6){\vector(1,0){0.8}}
\put(1.3,4.5){ $\nu_{\beta}$}
\put(2,4){\vector(1,0){0}}
\put(1.8,3.6){\vector(1,0){0.8}}
\put(2,3.1){$p-q$}
\put(2.5,4){\circle*{0.2}}
\multiput(2.5,4)(0.5,0){8}{\line(1,0){0.3}}
\put(4.65,4){\vector(1,0){0}}
\put(4.1,4.3){ $\tilde{\nu}_{X}$}
\put(4.1,3.6){\vector(1,0){0.8}}
\put(3.6,3.1){$k+p-q$}
\put(1.15,4){\circle*{0.2}}
\put(6.5,4){\circle*{0.2}}
\put(6.5,4){\line(1,0){2}}
\put(8,4.3){$\nu_{\alpha}$}
\put(6.8,3.6){\vector(1,0){0.8}}
\put(7.8,3.5){$p-q$}
\put(8,4){\vector(1,0){0}}
\put(4.5,4){
\qbezier(-2,0)(-2.01,0.35)(-1.88,0.68)
\qbezier(-1.88,0.68)(-1.78,1.03)(-1.53,1.29)
\qbezier(-1.53,1.29)(-1.3,1.55)(-1,1.73)
\qbezier(-1,1.73)(-0.69,1.9)(-0.35,1.97)
\qbezier(-0.35,1.97)(0,2.03)(0.35,1.97)
\qbezier(2,0)(2.01,0.35)(1.88,0.68)
\qbezier(1.88,0.68)(1.78,1.03)(1.53,1.29)
\qbezier(1.53,1.29)(1.3,1.55)(1,1.73)
\qbezier(1,1.73)(0.69,1.9)(0.35,1.97)
}
\put(4.8,6){\vector(1,0){0}}
\put(4.5,4){
\qbezier(-2,0)(-2.36,0.417)(-1.88,0.68)
\qbezier(-1.88,0.68)(-1.47,0.85)(-1.53,1.29)
\qbezier(-1.53,1.29)(-1.54,1.84)(-1,1.73)
\qbezier(-1,1.73)(-0.58,1.6)(-0.35,1.97)
\qbezier(-0.35,1.97)(0,2.35)(0.35,1.97)
\qbezier(2,0)(2.36,0.417)(1.88,0.68)
\qbezier(1.88,0.68)(1.47,0.85)(1.53,1.29)
\qbezier(1.53,1.29)(1.54,1.84)(1,1.73)
\qbezier(1,1.73)(0.58,1.6)(0.35,1.97)
}
\put(4.4,6.5){$\tilde{\chi}^{0}_{A}$}
\put(4.5,3.5){\qbezier(-0.35,1.97)(0,2.03)(0.35,1.97)}
\put(3.95,5.4){\vector(-4,-1){0}}
\put(4.4,5){$k$}
\rotatebox{30}{
\multiput(1.3,3.8)(0,-1){3}
{\qbezier(0,0)(-0.25,-0.25)(0,-0.5)
 \qbezier(0,-0.5)(0.25,-0.75)(0,-1.0)}
}
\put(0.8,2){$W$}
\put(-0.5,2.5){\vector(3,-4){0.6}}
\put(-0.9,2){$q$}
\end{picture}\vspace{-1.2cm}
\caption{One example of the diagrams in which we have to 
take into account the off-shellness for the neutrino propagation. 
In the neutrino propagator which is pointed out by the oval,
we must treat the neutrino as if it were a massless particle.}
\label{fig:Offshell}
\end{figure}
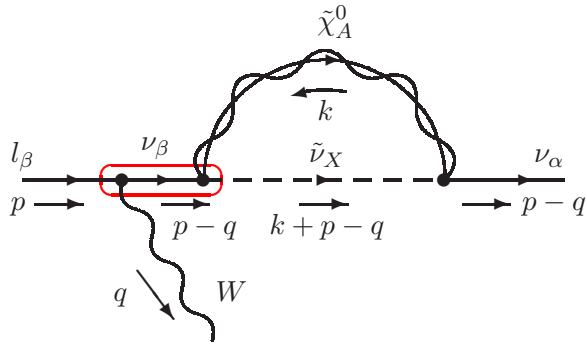

In the calculation of nLFV diagrams, 
we have to pay attention to the following two facts: 
(i) we regard neutrinos as highly off-shell particle, 
and 
(ii) we must avoid counting one contribution twice.
First, we explain the reason why neutrinos behave as
highly off-shell particles, 
and hence it is necessary to treat that neutrinos
are massless in diagrams for effective nLFV interactions.
In Fig. \ref{fig:Offshell}, from the viewpoint of the uncertainty principle,
the condition for neutrino oscillation to occur is described as\cite{FY}
\begin{eqnarray}
 \delta (p-q) \sim \frac{1}{\delta x}
  &\gg& \frac{1}{L} \sim\frac{\Delta m^2}{p-q},
\end{eqnarray}
where $\delta x$ is the uncertainty of the position
and the time in which neutrinos are produced, $\delta (p-q)$ is the
uncertainty of the energy-momentum of the outgoing neutrino, 
and $\Delta m^{2}$ is the neutrino mass squared difference corresponding
to the LBL oscillation experiment. 
The inequality results from the fact that 
the neutrino production position must be determined much more accurately 
than the baseline length $L$.  
The fact that the neutrino oscillation phenomena are observed in the LBL
experiment requires that the equality at the most right-hand side 
should be satisfied.
Thus, the uncertainty of the squared momentum must conform to the
following relation
\begin{eqnarray}
\delta \{ (p-q)^2 \} &\gg& \Delta m^2 \qquad \text{then} \qquad  
\langle (p-q)^2 \rangle  \gg m^2_{\nu},
\end{eqnarray}
where $\langle (p-q)^2 \rangle $ denotes the average of $(p-q)^2$.
This inequation shows that the average of the neutrino momentum is
much larger than its mass, and it follows from this that
neutrinos are generally highly off-shell fields in oscillation experiments.
This also means that all the diagrams for the nLFV interactions include the off-shell
neutrino as external lines. 
In the exact meaning of the field theory, we do not have the method to calculate
diagrams with off-shell external legs.
However, we can evaluate such diagrams by treating as if the neutrinos were massless.  
For the practical purpose, we make the following replacement which we
refer as the off-shell prescription 
in calculations of the type of diagram shown in
Fig. \ref{fig:Offshell};\footnote{%
The mass parameter $m_{\nu_{\alpha}}$ for the flavor eigenstate\cite{Giunti} 
appears in Eq. \eqref{eq:masslessNeutrino}. However, we finally neglect
it in our off-shell prescription.
}
\begin{align}
 \frac{(p-q)^2}{(p-q)^2-m_{\nu_\beta}^2} \Longrightarrow 
  \begin{cases}
   1,  &\hbox{(off-shell prescription)}, \\ 
   \frac{m_{\nu_\alpha}^2}{m_{\nu_\alpha}^2-m_{\nu_\beta}^2}, 
   & \hbox{(for usual on-shell particle case)}.
  \end{cases}
\label{eq:masslessNeutrino}
\end{align}

Next, we turn to the problem of the double counting and show the way to solve
it.
First, we note that 
we must calculate the process $\mu^{-} + I \rightarrow \tau^{-} + F$
with the method of the field theory, 
so that we must not calculate the nLFV effect for each stage because we cannot
observe the each stage separately. 
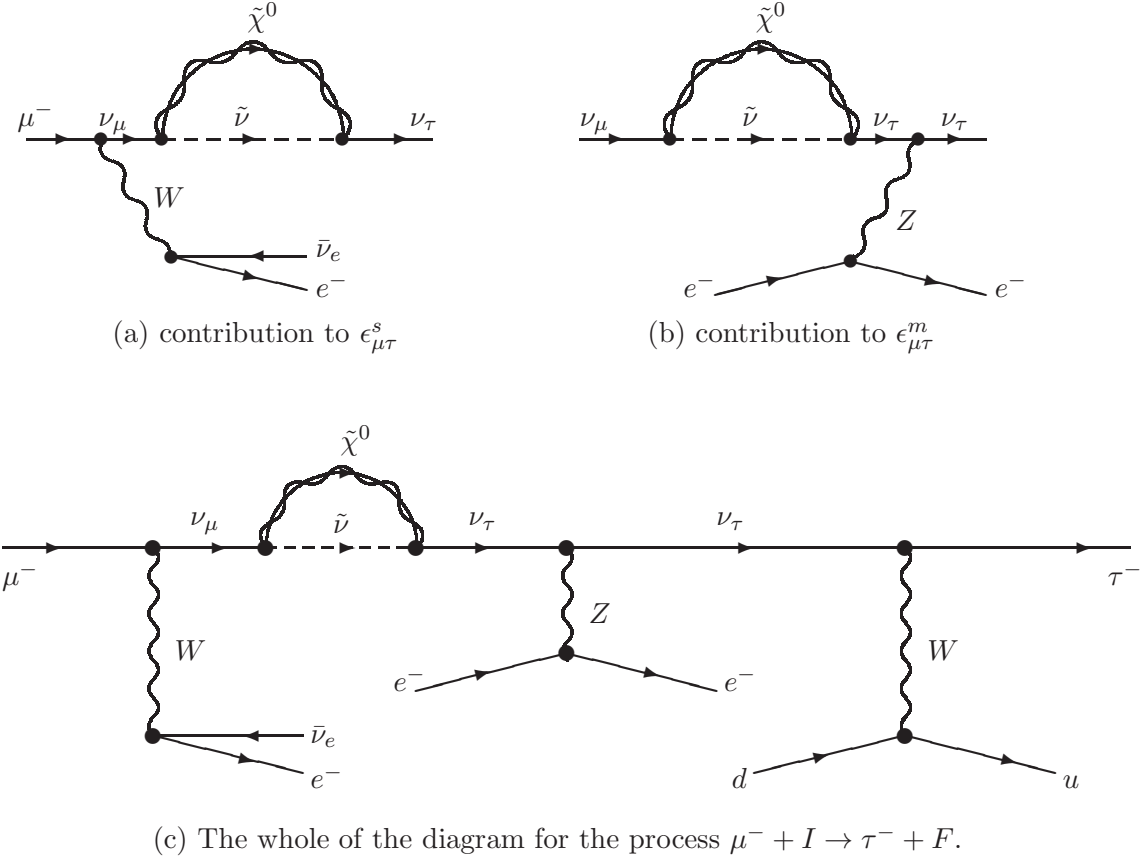
\begin{figure}[tb]
\unitlength=0.6cm
\begin{picture}(10,7)
\thicklines
\put(-0.5,4){\line(1,0){3}}
\put(0.5,4){\vector(1,0){0}}
\put(-0.7,4.3){$\mu^{-}$}
\put(1.1,4.3){$\nu_{\mu}$}
\put(2,4){\vector(1,0){0}}
\put(2.5,4){\circle*{0.3}}
\multiput(2.5,4)(0.5,0){8}{\line(1,0){0.3}}
\put(4.65,4){\vector(1,0){0}}
\put(4.1,4.3){$\tilde{\nu}$}
\put(1.15,4){\circle*{0.3}}
\put(6.5,4){\circle*{0.3}}
\put(6.5,4){\line(1,0){2}}
\put(8,4.3){$\nu_{\tau}$}
\put(8,4){\vector(1,0){0}}
\put(4.5,4){
\qbezier(-2,0)(-2.01,0.35)(-1.88,0.68)
\qbezier(-1.88,0.68)(-1.78,1.03)(-1.53,1.29)
\qbezier(-1.53,1.29)(-1.3,1.55)(-1,1.73)
\qbezier(-1,1.73)(-0.69,1.9)(-0.35,1.97)
\qbezier(-0.35,1.97)(0,2.03)(0.35,1.97)
\qbezier(2,0)(2.01,0.35)(1.88,0.68)
\qbezier(1.88,0.68)(1.78,1.03)(1.53,1.29)
\qbezier(1.53,1.29)(1.3,1.55)(1,1.73)
\qbezier(1,1.73)(0.69,1.9)(0.35,1.97)
}
\put(4.8,6){\vector(1,0){0}}
\put(4.5,4){
\qbezier(-2,0)(-2.36,0.417)(-1.88,0.68)
\qbezier(-1.88,0.68)(-1.47,0.85)(-1.53,1.29)
\qbezier(-1.53,1.29)(-1.54,1.84)(-1,1.73)
\qbezier(-1,1.73)(-0.58,1.6)(-0.35,1.97)
\qbezier(-0.35,1.97)(0,2.35)(0.35,1.97)
\qbezier(2,0)(2.36,0.417)(1.88,0.68)
\qbezier(1.88,0.68)(1.47,0.85)(1.53,1.29)
\qbezier(1.53,1.29)(1.54,1.84)(1,1.73)
\qbezier(1,1.73)(0.58,1.6)(0.35,1.97)
}
\put(4.4,6.5){$\tilde{\chi}^{0}$}
\rotatebox{30}{
\multiput(1.3,3.8)(0,-1){3}
{\qbezier(0,0)(-0.25,-0.25)(0,-0.5)
 \qbezier(0,-0.5)(0.25,-0.75)(0,-1.0)}
}
\put(0.4,2.5){$W$}
\put(0.8,1.4){\line(1,0){3}}
\put(0.8,1.4){\line(4,-1){3}}
\put(0.8,1.4){\circle*{0.3}}
\put(2.5,1.41){\vector(-1,0){0}}
\put(2.75,0.9){\vector(4,-1){0}}
\put(4,1.4){$\bar{\nu}_{e}$}
\put(4,0.5){$e^{-}$}
\put(-0.5,-0.5){(a) contribution to $\epsilon^{s}_{\mu\tau}$}
\end{picture}
\hspace{0.5cm}
\begin{picture}(10,7)
\thicklines
\put(0.5,4){\line(1,0){2}}
\put(1.5,4){\vector(1,0){0}}
\put(0.5,4.3){$\nu_{\mu}$}
\put(2.5,4){\circle*{0.3}}
\multiput(2.5,4)(0.5,0){8}{\line(1,0){0.3}}
\put(4.65,4){\vector(1,0){0}}
\put(4.1,4.3){$\tilde{\nu}$}
\put(7,4.3){$\nu_{\tau}$}
\put(8,4){\circle*{0.3}}
\put(6.5,4){\circle*{0.3}}
\put(6.5,4){\line(1,0){3}}
\put(9,4){\vector(1,0){0}}
\put(8.5,4.3){$\nu_{\tau}$}
\put(7.4,4){\vector(1,0){0}}
\put(4.5,4){
\qbezier(-2,0)(-2.01,0.35)(-1.88,0.68)
\qbezier(-1.88,0.68)(-1.78,1.03)(-1.53,1.29)
\qbezier(-1.53,1.29)(-1.3,1.55)(-1,1.73)
\qbezier(-1,1.73)(-0.69,1.9)(-0.35,1.97)
\qbezier(-0.35,1.97)(0,2.03)(0.35,1.97)
\qbezier(2,0)(2.01,0.35)(1.88,0.68)
\qbezier(1.88,0.68)(1.78,1.03)(1.53,1.29)
\qbezier(1.53,1.29)(1.3,1.55)(1,1.73)
\qbezier(1,1.73)(0.69,1.9)(0.35,1.97)
}
\put(4.8,6){\vector(1,0){0}}
\put(4.5,4){
\qbezier(-2,0)(-2.36,0.417)(-1.88,0.68)
\qbezier(-1.88,0.68)(-1.47,0.85)(-1.53,1.29)
\qbezier(-1.53,1.29)(-1.54,1.84)(-1,1.73)
\qbezier(-1,1.73)(-0.58,1.6)(-0.35,1.97)
\qbezier(-0.35,1.97)(0,2.35)(0.35,1.97)
\qbezier(2,0)(2.36,0.417)(1.88,0.68)
\qbezier(1.88,0.68)(1.47,0.85)(1.53,1.29)
\qbezier(1.53,1.29)(1.54,1.84)(1,1.73)
\qbezier(1,1.73)(0.58,1.6)(0.35,1.97)
}
\put(4.4,6.5){$\tilde{\chi}^{0}$}
\put(6.3,4){
\rotatebox{-30}{
\multiput(0,0)(0,-1){3}
{\qbezier(0,0)(-0.25,-0.25)(0,-0.5)
 \qbezier(0,-0.5)(0.25,-0.75)(0,-1.0)}}
}
\put(7.5,2){$Z$}
\put(6.5,1.3){\line(-4,-1){3}}
\put(6.5,1.3){\line(4,-1){3}}
\put(6.5,1.3){\circle*{0.3}}
\put(5,0.93){\vector(4,1){0}}
\put(8.5,0.8){\vector(4,-1){0}}
\put(2.8,0.5){$e^{-}$}
\put(9.7,0.5){$e^{-}$}
\put(2,-0.5){(b) contribution to $\epsilon^{m}_{\mu\tau}$}
\end{picture}\\
\unitlength=1cm
\begin{picture}(16,7)
\thicklines
\put(1,4){\line(1,0){3.5}}
\put(6.5,4){\line(1,0){9.5}}
\put(3,4){\circle*{0.2}}
\put(4.5,4){\circle*{0.2}}
\put(6.5,4){\circle*{0.2}}
\put(13,4){\circle*{0.2}}
\put(8.5,4){\circle*{0.2}}
\multiput(3,4)(0,-0.5){5}
{\qbezier(0,0)(0.125,-0.125)(0,-0.25)
 \qbezier(0,-0.25)(-0.125,-0.375)(0,-0.5)}
\multiput(8.5,4)(0,-0.5){3}
{\qbezier(0,0)(0.125,-0.125)(0,-0.25)
 \qbezier(0,-0.25)(-0.125,-0.375)(0,-0.5)}
\multiput(13,4)(0,-0.5){5}
{\qbezier(0,0)(0.125,-0.125)(0,-0.25)
 \qbezier(0,-0.25)(-0.125,-0.375)(0,-0.5)}
\put(3,1.5){\line(1,0){2}}
\put(3,1.5){\line(4,-1){2}}
\put(3,1.5){\circle*{0.2}}
\put(8.5,2.6){\line(-4,-1){2}}
\put(8.5,2.6){\line(4,-1){2}}
\put(8.5,2.6){\circle*{0.2}}
\put(13,1.5){\line(-4,-1){2}}
\put(13,1.5){\line(4,-1){2}}
\put(13,1.5){\circle*{0.2}}
\put(1,3.5){$\mu^{-}$}
\put(15.7,3.5){$\tau^{-}$}
\put(3.3,2.5){$W$}
\put(13.3,2.5){$W$}
\put(8.8,3){$Z$}
\put(5.1,1.4){$\bar{\nu}_{e}$}
\put(5.1,0.8){$e^{-}$}
\put(6.2,2.1){$e^{-}$}
\put(10.6,2.1){$e^{-}$}
\put(10.7,0.8){$d$}
\put(15.1,0.8){$u$}
\put(3.5,4.3){$\nu_{\mu}$}
\put(7.2,4.3){$\nu_{\tau}$}
\put(10.5,4.3){$\nu_{\tau}$}
\put(5.5,5.3){$\tilde{\chi}^{0}$}
\put(5.4,4.2){$\tilde{\nu}$}
\put(1.8,4){\vector(1,0){0}}
\put(4,4){\vector(1,0){0}}
\put(5.7,4){\vector(1,0){0}}
\put(7.5,4){\vector(1,0){0}}
\put(11,4){\vector(1,0){0}}
\put(15.5,4){\vector(1,0){0}}
\put(4.2,1.5){\vector(-1,0){0}}
\put(4.36,1.15){\vector(4,-1){0}}
\put(7.5,2.35){\vector(4,1){0}}
\put(9.75,2.3){\vector(4,-1){0}}
\put(12,1.25){\vector(4,1){0}}
\put(14.5,1.13){\vector(4,-1){0}}
\put(5.5,4){
\qbezier(-1,0)(-1.005,0.175)(-0.945,0.34)
\qbezier(-0.945,0.34)(-0.89,0.5015)(-0.765,0.645)
\qbezier(-0.765,0.645)(-0.65,0.775)(-0.5,0.865)
\qbezier(-0.5,0.865)(-0.345,0.95)(-0.175,0.985)
\qbezier(-0.175,0.985)(0,1.015)(0.175,0.985)
\qbezier(1,0)(1.005,0.175)(0.945,0.34)
\qbezier(0.945,0.34)(0.89,0.5015)(0.765,0.645)
\qbezier(0.765,0.645)(0.65,0.775)(0.5,0.865)
\qbezier(0.5,0.865)(0.345,0.95)(0.175,0.985)
}
\put(5.7,5){\vector(1,0){0}}
\put(5.5,4){
\qbezier(-1,0)(-1.18,0.2085)(-0.94,0.34)
\qbezier(-0.94,0.34)(-0.735,0.425)(-0.765,0.645)
\qbezier(-0.765,0.645)(-0.77,0.92)(-0.5,0.865)
\qbezier(-0.5,0.865)(-0.29,0.8)(-0.175,0.985)
\qbezier(-0.175,0.985)(0,1.175)(0.175,0.985)
\qbezier(1,0)(1.18,0.2085)(0.94,0.34)
\qbezier(0.94,0.34)(0.735,0.425)(0.765,0.645)
\qbezier(0.765,0.645)(0.77,0.92)(0.5,0.865)
\qbezier(0.5,0.865)(0.29,0.8)(0.175,0.985)
}
\multiput(4.5,4)(0.25,0){8}{\line(1,0){0.15}}
\put(3,0){(c) The whole of the diagram for the process $\mu^{-} + I \rightarrow \tau^{-} +F$.}
\end{picture}
\caption{Schematic explanation for the double counting problem.
If we count the diagram-(a) into $\epsilon^{s}_{\mu\tau}$ and we also do
 the diagram-(b) into $\epsilon^{m}_{\mu\tau}$, then we count the
 diagram-(c) twice.}
\label{fig:WholeDiagram}
\end{figure}
If we calculate the diagrams shown in Figs. \ref{fig:WholeDiagram}-(a)
for $\epsilon^{s}_{\mu\tau}$
and  \ref{fig:WholeDiagram}-(b) for $\epsilon^m_{\mu\tau}$, 
then it means that we count twice the diagram of
Fig. \ref{fig:WholeDiagram}-(c)
into the calculation of the process $\mu^{-} +I \rightarrow \tau^{-} +F$. 
In order to avoid doubly counting, we have to get rid of one of them.
For example, we should not include the contribution of the diagram
of Fig. \ref{fig:WholeDiagram}-(b) in $\epsilon^m_{\mu\tau}$.  
A similar situation occurs among $\epsilon^d_{\mu\tau}$ and $\epsilon^{m}_{\mu\tau}$.
The penguin contribution to $\epsilon^d_{\mu\tau}$ is essentially given 
by the complex conjugate of that to the corresponding $\epsilon^s_{\tau \mu}$. 
However, we must eliminate the contribution from diagrams similar to 
Fig. \ref{fig:WholeDiagram}-(a) from $\epsilon^{d}_{\mu\tau}$.

Finally, we should notice another thing that is important for the double
counting problem;
Which stage ({\it source} or {\it matter}) 
does the contribution of Fig. \ref{fig:WholeDiagram}-(c)
belong to?
For example, instead of removing the contribution of 
Fig. \ref{fig:WholeDiagram}-(b) in $\epsilon^m_{\mu\tau}$, 
we can eliminate the contribution of 
Fig. \ref{fig:WholeDiagram}-(a) 
in $\epsilon^s_{\mu\tau}$. 
For this ambiguity, we adopt the way to divide the diagram into
each stage so that epsilon parameters in each stage will disappear in
the limit where $m_{\rm S} \rightarrow \infty$ after using the off-shell
prescription.
The $SU(2)_{L}$ symmetry is recovered and
then $\epsilon^{s,m,d}_{\mu\tau}$ should disappear in the large SUSY scale limit.
This is shown analytically.    
Note that each diagram gives rather large contribution and stays almost
constant in the large SUSY scale limit. 
The cancellation among the
diagrams is highly nontrivial. Therefore,
by checking the cancellation among the diagrams, 
we can be confident about the legitimacy of our treatment 
for the internal neutrino lines.

\subsection{Numerical Study}

\begin{figure}[htb]
\unitlength=1cm
\begin{picture}(5,5.5)
\includegraphics[width=5cm]{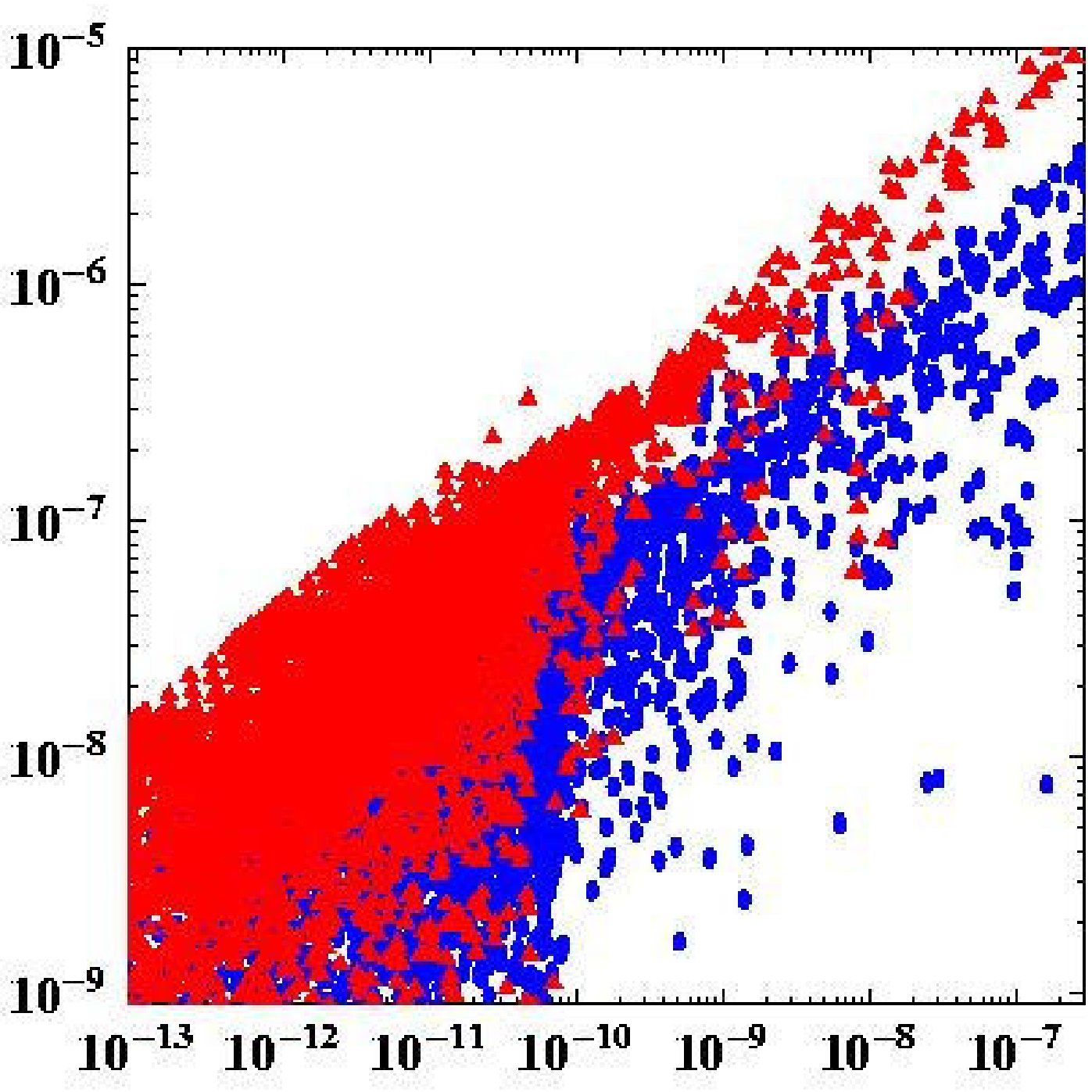}
\put(-4,4.2){$\epsilon^{s}_{\mu\tau}$}
\put(-5.5,2.8){$|\epsilon|$}
\put(-3.3,-0.3){Br$(\tau\rightarrow \mu \gamma)$}
\end{picture}
\begin{picture}(5,5.5)
\includegraphics[width=5cm]{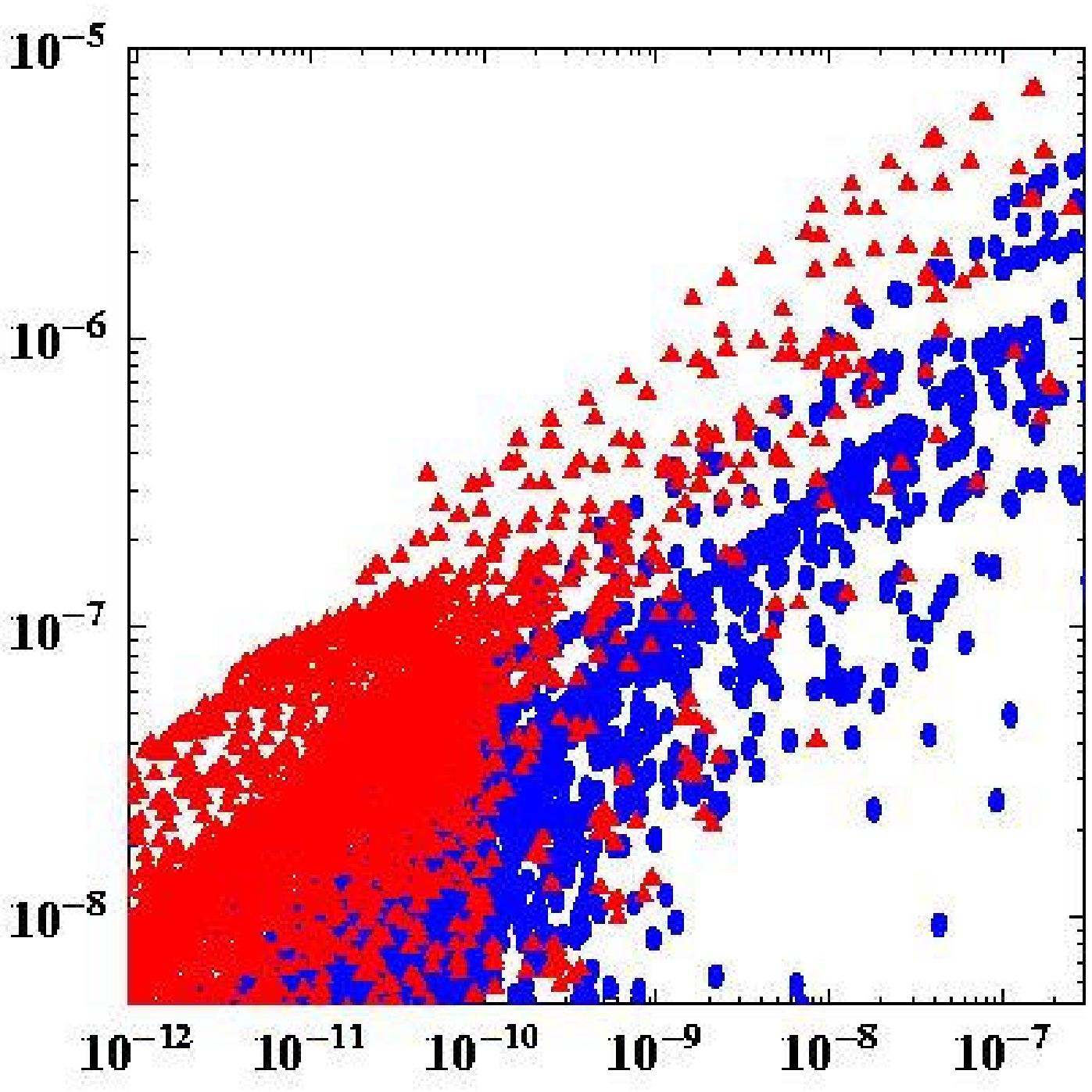}
\put(-4,4.2){$\epsilon^{m}_{\mu\tau}$}
\put(-3.3,-0.3){Br$(\tau\rightarrow \mu \gamma)$}
\end{picture}
\begin{picture}(5,5.5)
\includegraphics[width=5cm]{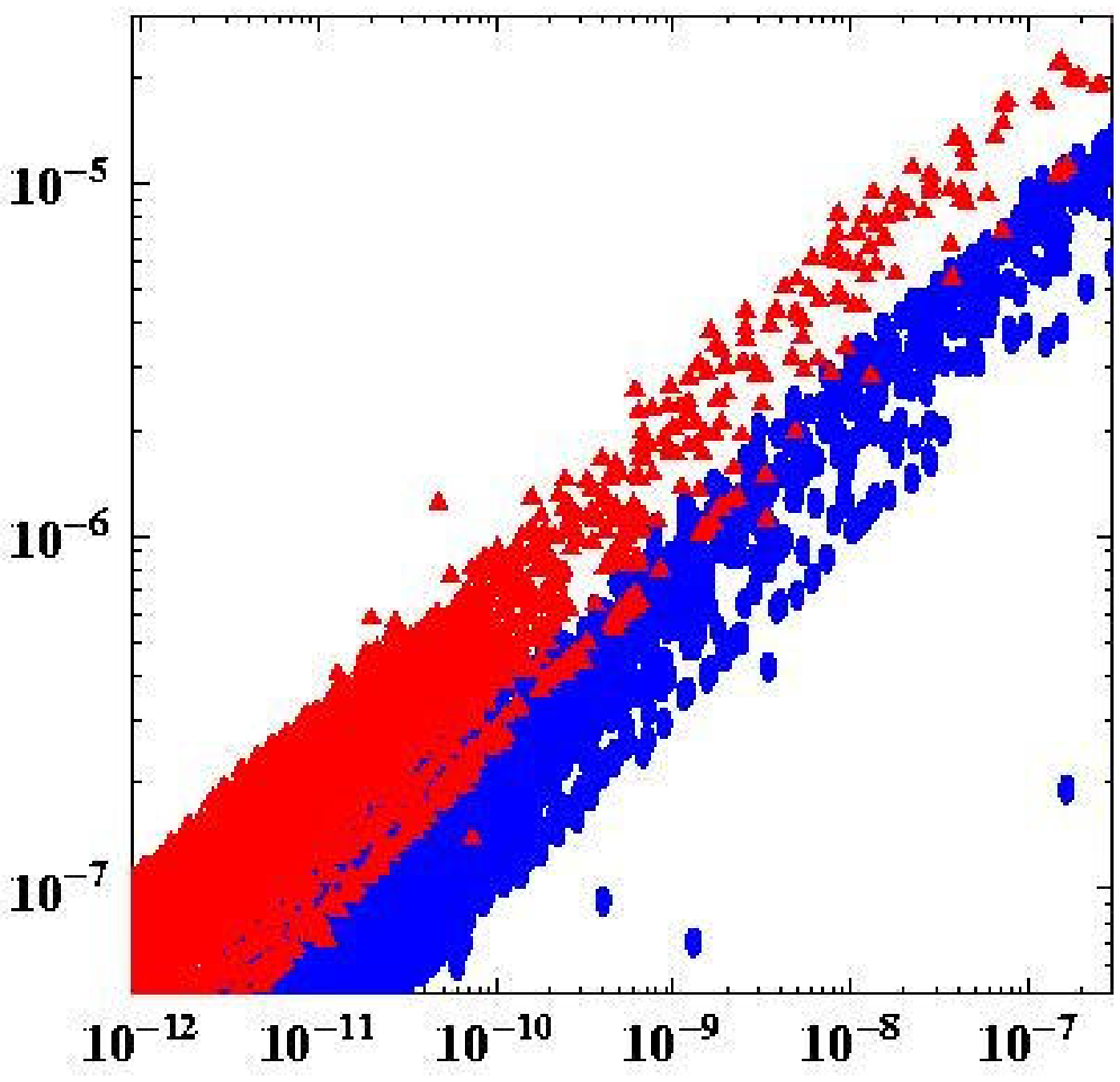}
\put(-4,4.2){$\epsilon^{d}_{\mu\tau}$}
\put(-3.3,-0.3){Br$(\tau\rightarrow \mu \gamma)$}
\end{picture}
\caption{Scatter plots for size of nLFV parameters $\epsilon^{s}_{\mu\tau}$,
 $\epsilon^{m}_{\mu\tau}$ and $\epsilon^{d}_{\mu\tau}$.
Dots are for $\tan\beta=10$ and triangles are for $\tan\beta=3$.
We here fix the soft SUSY breaking parameters as $a_{0} = 0$ and $\mu > 0$.}
\label{fig:scatter-plots} 
\end{figure}

A numerical study to evaluate the epsilon parameters is necessary in order to make it
clear whether our naive estimation
Eq. \eqref{eq:naive-estimation-epsilon} 
is correct. 
We here assume the universal soft SUSY
breaking at $M_{G}$ 
and adopt the scenario of the radiative electroweak symmetry
breaking\cite{IKKT}. 
The detail of calculations is shown in Appendices.
We scan the values for the soft SUSY
breaking parameters $m_{0}$ and $M_{1/2}$ at the range of 100 - 1,000 GeV,
and also scan the elements of the neutrino Yukawa matrix.
We here take the normal hierarchy for the neutrino mass matrix.
The scatter plots for size of $\epsilon^{s}_{\mu\tau}$,
$\epsilon^{m}_{\mu\tau}$ and $\epsilon^{d}_{\mu\tau}$ are presented in
Fig. \ref{fig:scatter-plots}.
All of them take similar behavior.
As expected from Eq. \eqref{eq:naive-estimation-epsilon},
the nLFV parameters correlate with the branching ratio of 
the process $\tau\rightarrow\mu\gamma$.
However, the size of the parameters is much smaller than that of the
naive estimation.
This fact can be understood by the cancellation among the diagrams which
contribute to the nLFV interaction.
In the $SU(2)_{L}$ symmetric limit,
the diagrams for the penguin contribution to the nLFV interaction 
must cancel each other out because of the gauge symmetry.
Since the diagrams for nLFV interactions are induced at the $m_{\rm S}$ scale,
in the limit where $m_{\rm S} \gg m_{Z}$, the $SU(2)_{L}$
symmetry is assumed to be recovered. 
Even in the case where $m_{\rm S} \sim \mathcal{O}(100)$ GeV,
the cancellation is rather significant. 
Thus, our naive estimation must be modified so that
the additional suppression factor $m_{Z} / m_{\rm S}$ is
introduced.
It may be worth pointing out that the epsilon parameters do not 
strongly depend on the value of $\tan\beta$ unlike the branching ratio
of the cLFV process.
This fact arises from the difference in the structure of the
chirality in each process.
The process $\tau \rightarrow \mu \gamma$ is dominated by the diagrams
including the left-right mixing of the slepton which is proportional
to $\tan\beta$.
The nLFV processes do not pick up such the left-right mixing term 
because a chirality flip is not necessary.
Therefore, nLFV search may be advantageous
in the case where $\tan\beta$ is relatively small.
We also note that
it is obvious from Eq. \eqref{eq:RGESolution} that 
both nLFV and cLFV are enhanced when $a_{0}$ takes a large value.  

\section{Summary}
\label{sec:summary}

We summarize our study and give some discussions. 
It is known that magnitude of nLFV couplings 
can become large enough to be detected at future LBL experiments 
within model-independent approach.
We here considered the nLFV interactions
in the MSSMRN under the universal soft SUSY breaking scenario
which is one of the most promising candidates for the physics
beyond the SM.

We find that in this scenario the nLFV couplings cannot be significant, 
and hence it is quite difficult to observe these effects at future
oscillation experiments.
The reason why they are strongly suppressed is that the 
$SU(2)_{L}$ gauge symmetry is approximately maintained. 
All the particles and the interactions which can
generate nLFV interactions respect the $SU(2)_{L}$ symmetry 
in the limit $m_S\rightarrow\infty$. 
Although each diagram to contribute to nLFV interactions can become large, 
a brilliant cancellation among those diagrams occurs. 
Therefore, the penguin contributions are strongly suppressed.

We adopt the approximation for the calculation of penguin diagrams 
which is explained in Sec. \ref{sec:analytic-calc}.
In order to confirm the validity of this approximation,
we need to make the calculation for the process shown in
Fig. \ref{fig:WholeDiagram}-(c)
by using the method of the field theory.
However, the calculation which we adopt here must be 
reliable in the sense of the field theory
because the consistencies, 
i.e., the recovery of $SU(2)_{L}$
gauge symmetry, are obviously maintained in our calculation.

Finally, we mention our future work.
Since the decay process of a muon differs from that of a pion, 
we might expect that sizable new physics effect exists only 
in the decay of pion.
Thus, it is necessary to investigate nLFV effects in the MSSMRN at 
superbeam experiments individually.
Furthermore, because within the model independent approach, 
epsilon parameters can be still significant
and there are lots of the other models than the constrained MSSMRN
which can explain the neutrino masses and
the lepton mixings, we need to examine such possibilities.

\vspace*{1cm}
\noindent
{\large \it Acknowledgments}\\
J. S. thanks J. W. F. Valle for useful discussion.
The work of T. O. is supported in part by Japan Society for Promotion of
Science No. 3693. 
J. S. was supported by the Grant-in-Aid for Scientific Research
on Priority Area No.16038202 and 14740168.

\appendix

\section{Model}
\label{app:Model}

We basically follow the notation which is adopted in Ref. \cite{HMTYY}.  
However,
we pay attention to the fact that the mixing matrices to diagonalize
the mass matrices for sferimons, chargino and neutralino are complex
matrices, in general.

The mass matrices for charged slepton, down-type squark and up-type squark
are
\begin{align}
-\mathcal{L}_{\tilde{f}} =&
\begin{pmatrix}
\tilde{f}_{L}^{*} & \tilde{f}_{R}^{*}
\end{pmatrix}^{\alpha}
{\left( M_{\tilde{f}}^{2} \right)_{\alpha}}^{\beta}
\begin{pmatrix}
\tilde{f}_{L} \\ 
\tilde{f}_{R}
\end{pmatrix}_{\beta}
=
\begin{pmatrix}
\tilde{f}_{L}^{*} & \tilde{f}_{R}^{*}
\end{pmatrix}
\begin{pmatrix}
M_{\tilde{f}_{LL}}^{2} 
& (M_{\tilde{f}_{LR}}^{2})^{\dagger} \\
M_{\tilde{f}_{LR}}^{2}
&M_{\tilde{f}_{RR}}^{2}
\end{pmatrix}
\begin{pmatrix}
\tilde{f}_{L} \\ 
\tilde{f}_{R}
\end{pmatrix},
\end{align}
where
\begin{align}
M_{\tilde{f}_{LL}}^{2} 
 &= 
 \begin{cases}
  m_{\tilde{f}_{L}}^{2}
   +
   \left(f_{f}^{\dagger} f_{f} \right) \frac{v^{2}}{\sqrt{2}} \sin^{2} \beta
   +
   m_{Z}^{2} \cos 2 \beta \left\{ \left(\frac{1}{2}\right) - Q^{f}_{\rm em} \sin^{2}
 \theta_{W} \right\}, \qquad \text{($f=u$)},\\
  m_{\tilde{f}_{L}}^{2}
   +
   \left(f_{f}^{\dagger} f_{f} \right) \frac{v^{2}}{\sqrt{2}} \cos^{2} \beta
   +
   m_{Z}^{2} \cos 2 \beta \left\{ \left( - \frac{1}{2} \right) - Q^{f}_{\rm em} \sin^{2}
 \theta_{W} \right\}, \quad \text{($f=l,d$)},
 \end{cases}\\
M_{\tilde{f}_{RR}}^{2}
 &=
 \begin{cases}
  m_{\tilde{f}_{R}}^{2}
   +
   \left(f_{f} f_{f}^{\dagger} \right) \frac{v^{2}}{\sqrt{2}} \sin^{2} \beta
   -
   m_{Z}^{2} \cos 2 \beta \left\{- Q^{f}_{\rm em} \sin^{2}
 \theta_{W} \right\}, \qquad \text{($f=u$)},\\
  m_{\tilde{f}_{R}}^{2}
   +
   \left(f_{f} f_{f}^{\dagger} \right) \frac{v^{2}}{\sqrt{2}} \cos^{2} \beta
   -
   m_{Z}^{2} \cos 2 \beta \left\{- Q^{f}_{\rm em} \sin^{2}
 \theta_{W} \right\},  \qquad \text{($f=l,d$)},
 \end{cases}\\
M_{\tilde{f}_{LR}}
 &=
\begin{cases}
 -A_{f} \frac{v}{\sqrt{2}} \sin\beta 
 - \mu f_{f} \frac{v}{\sqrt{2}} \sin\beta
   \cot \beta, \qquad \text{($f=u$)}, \\
 A_{f} \frac{v}{\sqrt{2}} \cos\beta 
 - \mu f_{f} \frac{v}{\sqrt{2}} \cos\beta
   \tan \beta, \qquad \text{($f=l,d$)}.
\end{cases}
\end{align}
Here,
the indices $\alpha$ and $\beta$ are for interaction eigenstates for
thier superpartner fermion fields.
We take the basis where the mass matrix for the charged lepton field is
diagonal,
so that the index for chraged lepton indicates its interaction eigenstate
and its mass eigenstate, simultaneously. 
The unitary matrix $U_{\tilde{f}}$ is defined as
\begin{align}
{(U_{\tilde{f}})_{X}}^{\alpha}
 {(M_{\tilde{f}}^{2})_{\alpha}}^{\beta}
 {(U_{\tilde{f}}^{\dagger})_{\beta}}^{Y}
 =
 \text{diag}
 (m_{\tilde{f}_{X}}^{2}) {\delta_{X}}^{Y}.
\end{align}
The relations between the mass eigenstates and the interaction
eigenstates are 
\begin{align}
 \tilde{f}_{X} = {(U_{\tilde{f}})_{X}}^{\alpha} \tilde{f}_{L \alpha}+
  {(U_{\tilde{f}})_{X}}^{\alpha+3} \tilde{f}_{R \alpha}, 
\end{align}
\begin{align}
\tilde{f}_{L \alpha} = {(U_{\tilde{f}}^{\dagger})_{\alpha}}^{X} \tilde{f}_{X},
 \qquad
\tilde{f}_{R \alpha} = {(U_{\tilde{f}}^{\dagger})_{\alpha+3}}^{X} \tilde{f}_{X}.
\label{eq:sl-flavor-mass}
\end{align}

The sneutrino mass term is also given as
\begin{align}
-\mathcal{L}_{\tilde{\nu}} =
 \tilde{\nu}^{* \alpha} 
 {(M_{\tilde{\nu}}^{2})_{\alpha}}^{\beta}
 \tilde{\nu}_{\beta},
\end{align}
where
\begin{align}
(M_{\tilde{\nu}}^{2})
=
m_{\tilde{L}}^{2}
   +
  m_{\nu}^{\dagger} m_{\nu}
   +
  m_{Z}^{2} \cos 2 \beta \left( \frac{1}{2} \right),
\end{align}
where $m_{\nu}$ is the neutrino mass matrix which is induced by the
seesaw mechanism;
\begin{align}
m_{\nu} = f_{\nu}^{\sf T} M^{-1} f_{\nu} \frac{v^{2}}{2} \sin \beta.
\end{align}
The unitary matrix $U_{\tilde{\nu}}$ is defined as
\begin{align}
{(U_{\tilde{\nu}})_{X}}^{\alpha}
 {(M_{\tilde{\nu}}^{2})_{\alpha}}^{\beta}
 {(U_{\tilde{\nu}}^{\dagger})_{\beta}}^{Y}
 =
 \text{diag}
 (m_{\tilde{\nu}_{X}}^{2}) {\delta_{X}}^{Y}.
\end{align}
The relations between the mass eigenstates and the interaction
eigenstates are 
\begin{align}
 \tilde{\nu}_{X} = {(U_{\tilde{\nu}})_{X}}^{\alpha} \tilde{\nu}_{L \alpha}, \qquad
\tilde{\nu}_{L \alpha} = {(U_{\tilde{\nu}}^{\dagger})_{\alpha}}^{X}
 \tilde{\nu}_{X}.
\label{eq:snu-flavor-mass}
\end{align}


The chargino mass term in the 2-spinor representation is 
\begin{align}
-\mathcal{L}_{\tilde{\chi}^{-}}
=
\begin{pmatrix}
  \tilde{w}^{+} & \tilde{h}_{u}^{+}
 \end{pmatrix}^{i}
{\left(M_{C}\right)_{i}}^{j}
  \begin{pmatrix}
  \tilde{w}^{-} \\ 
   \tilde{h}_{d}^{-}
 \end{pmatrix}_{j}
=
 \begin{pmatrix}
  \tilde{w}^{+} & \tilde{h}_{u}^{+}
 \end{pmatrix}
 \begin{pmatrix}
  M_{2} & \sqrt{2} m_{W} \cos\beta \\
  \sqrt{2} m_{W} \sin\beta &\mu
 \end{pmatrix}
  \begin{pmatrix}
  \tilde{w}^{-} \\ 
   \tilde{h}_{d}^{-}
 \end{pmatrix}.
\end{align}
The diagonalization is done by unitary matrices $U_{R}$ and $U_{L}$ as
\begin{align}
{(U_{R})_{A}}^{i} {(M_{C})_{i}}^{j} {(U_{L}^{\dagger})_{j}}^{B} 
=\text{diag}(M_{\tilde{\chi}^{-}_{A}}) {\delta_{A}}^{B}, 
\end{align}
The relations between the interaction eigenstates and the mass
eigenstates are
\begin{align}
 (\tilde{x}^{-}_{A})_{a} = {(U_{L})_{A}}^{i}
 \begin{pmatrix}
  (\tilde{w}^{-})_{a} \\ 
  (\tilde{h}_{d}^{-})_{a}
 \end{pmatrix}_{i}, \qquad
 (\tilde{x}^{+A})_{a} = 
 {(U_{R}^{*})^{A}}_{i}
 \begin{pmatrix}
  (\tilde{w}^{+})_{a} \\
  (\tilde{h}_{u}^{+})_{a}
 \end{pmatrix}^{i},
\end{align}
where $a$ ($\dot{a}$) denotes index for the Lorentz spinor for ${\bf 2}$
($\bf{\bar{2}}$) under $SL(2,C)$.
We here adopt the same rule as that in Ref. \cite{HaberKane}.  
The 4-spinors for mass eigenstates are constructed as 
\begin{align}
\tilde{\chi}^{-}_{A} = 
 \begin{pmatrix}
  (\tilde{x}^{-}_{A})_{a} \\
  (\overline{\tilde{x}^{+}}_{A})^{\dot{a}}
 \end{pmatrix}, \qquad 
\tilde{\chi}^{+ A} = 
 \begin{pmatrix}
  (\tilde{x}^{+ A})_{a} \\
  (\overline{\tilde{x}^{-}}^{A})^{\dot{a}}
 \end{pmatrix},
\end{align}
and then those for interaction eigenstates are
\begin{align}
 \tilde{W}^{-} &= 
 \begin{pmatrix}
  (\tilde{w}^{-})_{a} \\
  (\overline{\tilde{w}^{+}})^{\dot{a}}
 \end{pmatrix}
 =
 \begin{pmatrix}
  {(U_{L}^{\dagger})_{1}}^{A} (\tilde{x}^{-}_{A})_{a} \\
  {(U_{R}^{\dagger})_{1}}^{A} (\overline{\tilde{x}^{+}}_{A})^{\dot{a}}
 \end{pmatrix}, 
 \qquad
 \tilde{W}^{+} = 
 \begin{pmatrix}
  (\tilde{w}^{+})_{a} \\
  (\overline{\tilde{w}^{-}})^{\dot{a}}
 \end{pmatrix}
 =
 \begin{pmatrix}
  {(U_{R}^{\sf T})^{1}}_{A} (\tilde{x}^{+ A})_{a} \\
  {(U_{L}^{\sf T})^{1}}_{A} (\overline{\tilde{x}^{-}}^{A})^{\dot{a}}
 \end{pmatrix},  \nonumber \\
\tilde{H}^{-} &=
 \begin{pmatrix}
  (\tilde{h}_{d}^{-})_{a} \\
  (\overline{\tilde{h}_{u}^{+}})^{\dot{a}}
 \end{pmatrix}
  =
 \begin{pmatrix}
  {(U_{L}^{\dagger})_{2}}^{A} (\tilde{x}^{-}_{A})_{a} \\
  {(U_{R}^{\dagger})_{2}}^{A} (\overline{\tilde{x}^{+}}_{A})^{\dot{a}}
 \end{pmatrix}
 , \qquad
\tilde{H}^{+} 
 =
 \begin{pmatrix}
  (\tilde{h}_{u}^{+})_{a} \\
  (\overline{\tilde{h}_{d}^{-}})^{\dot{a}}
 \end{pmatrix}
 =
 \begin{pmatrix}
  {(U_{R}^{\sf T})^{2}}_{A} (\tilde{x}^{+ A})_{a} \\
  {(U_{L}^{\sf T})^{2}}_{A} (\overline{\tilde{x}^{-}}^{A})^{\dot{a}}
 \end{pmatrix}.
\end{align}

The neutralino mass term in 2-spinor representation is
\begin{align}
-\mathcal{L}_{\tilde{\chi}^{0}} =&
\frac{1}{2}
\begin{pmatrix}
 \tilde{b}^{0} & \tilde{w}^{0} & \tilde{h}^{0}_{d} & \tilde{h}^{0}_{u}
\end{pmatrix}_{i}
(M_{N})^{ij}
\begin{pmatrix}
 \tilde{b}^{0} \\ \tilde{w}^{0} \\ \tilde{h}^{0}_{d} \\ \tilde{h}^{0}_{u}
\end{pmatrix}_{j} \nonumber \\
=&
\frac{1}{2}
\begin{pmatrix}
 \tilde{b}^{0} & \tilde{w}^{0} & \tilde{h}^{0}_{d} & \tilde{h}^{0}_{u}
\end{pmatrix} \\
&\times
\begin{pmatrix}
M_{1} & 0 & -m_{Z} \sin \theta_{W} \cos\beta & m_{Z} \sin \theta_{W}
 \sin\beta \\
0 & M_{2} & m_{Z} \cos \theta_{W} \cos\beta &  -m_{Z} \cos \theta_{W}
 \sin\beta \\
 -m_{Z} \sin \theta_{W} \cos\beta &  m_{Z} \cos \theta_{W} \cos\beta
& 0 & -\mu \\
 m_{Z} \sin \theta_{W} \sin\beta &  -m_{Z} \cos \theta_{W} \sin\beta
& -\mu & 0 
\end{pmatrix}
\begin{pmatrix}
 \tilde{b}^{0} \\ \tilde{w}^{0} \\ \tilde{h}^{0}_{d} \\ \tilde{h}^{0}_{u}
\end{pmatrix}. \nonumber 
\end{align}
The unitary matrix $U_{N}$ is defined as
\begin{align}
{(U_{N}^{*})^{A}}_{i} 
 (M_{N})^{ij} 
 {(U_{N}^{\dagger})_{j}}^{B} 
= \text{diag}(M_{\tilde{\chi}^{0}_{A}}) \delta^{AB}.
\end{align}
The relations between the interaction eigenstates and the mass
eigenstates are
\begin{align}
\tilde{x}^{0}_{A} = {(U_{N})_{A}}^{i}
\begin{pmatrix}
 \tilde{b}^{0} \\ \tilde{w}^{0} \\ \tilde{h}^{0}_{d} \\ \tilde{h}^{0}_{u}
\end{pmatrix}_{i}  
\end{align}
Using this 2-spinor $\tilde{x}^{0}$, the 4-Majorana spinor can be
constructed as
\begin{align}
{\rm P}_{L} \tilde{\chi}^{0}_{A}
 +
{\rm P}_{R} \tilde{\chi}^{0 A}
 =
  \begin{pmatrix}
   (x^{0}_{A})_{a} \\
   (\overline{x^{0}}^{A})^{\dot{a}}
  \end{pmatrix}.
\end{align}
%
The interaction eigenstates are
\begin{align}
\tilde{B}^{0}
 =\begin{pmatrix}
   (\tilde{b}^{0})_{a} \\
   (\overline{\tilde{b}^{0}})^{\dot{a}}
  \end{pmatrix}
 =
 \begin{pmatrix}
  {(U_{N}^{\dagger})_{1}}^{A} (\tilde{x}^{0}_{A})_{a}\\
  {(U_{N}^{\sf T})^{1}}_{A} (\overline{\tilde{x}^{0}}^{A})^{\dot{a}}
 \end{pmatrix},\qquad 
\tilde{W}^{0}
 =\begin{pmatrix}
   (\tilde{w}^{0})_{a} \\
   (\overline{\tilde{w}^{0}})^{\dot{a}}
  \end{pmatrix}
 =
 \begin{pmatrix}
  {(U_{N}^{\dagger})_{2}}^{A} (\tilde{x}^{0}_{A})_{a}\\
  {(U_{N}^{\sf T})^{2}}_{A} (\overline{\tilde{x}^{0}}^{A})^{\dot{a}}
 \end{pmatrix}, \nonumber \\
\tilde{H}^{0}_{d}
 =\begin{pmatrix}
   (\tilde{h}^{0}_{d})_{a} \\
   (\overline{\tilde{h}^{0}_{d}})^{\dot{a}}
  \end{pmatrix}
 =
 \begin{pmatrix}
  {(U_{N}^{\dagger})_{3}}^{A} (\tilde{x}^{0}_{A})_{a}\\
  {(U_{N}^{\sf T})^{3}}_{A} (\overline{\tilde{x}^{0}}^{A})^{\dot{a}}
 \end{pmatrix},\qquad 
\tilde{H}^{0}_{u}
 =\begin{pmatrix}
   (\tilde{h}^{0}_{u})_{a} \\
   (\overline{\tilde{h}^{0}_{u}})^{\dot{a}}
  \end{pmatrix}
 =
 \begin{pmatrix}
  {(U_{N}^{\dagger})_{4}}^{A} (\tilde{x}^{0}_{A})_{a}\\
  {(U_{N}^{\sf T})^{4}}_{A} (\overline{\tilde{x}^{0}}^{A})^{\dot{a}}
 \end{pmatrix}.
\end{align}

The Lagrangian for gaugino-sfermion-fermion interactions is described as
\begin{align}
\mathcal{L}_{\text{int}} =&
 \bar{l}^{\alpha} 
 \left\{
 (C^{(l)}_{R})_{\alpha}^{AX} {\rm P}_{R} 
 +
 (C^{(l)}_{L})_{\alpha}^{AX} {\rm P}_{L} 
 \right\}
 \tilde{\chi}^{-}_{A} \tilde{\nu}_{X}
 +
 \bar{\nu}^{\alpha} 
 (C^{(\nu)}_{R})_{\alpha A}^{X} {\rm P}_{R} 
 \tilde{\chi}^{+ A} \tilde{l}_{X} \nonumber \\
&+
 \bar{l}^{\alpha} 
 (N^{(l)}_{R})_{\alpha A}^{X} {\rm P}_{R}
 \tilde{\chi}^{0 A}
 \tilde{l}_{X}
 +
 \bar{l}^{\alpha} 
 (N^{(l)}_{L})_{\alpha}^{AX} {\rm P}_{L}
 \tilde{\chi}^{0}_{A}
 \tilde{l}_{X}
+
\bar{\nu}^{\alpha} 
 (N^{(\nu)}_{R})_{\alpha A}^{X} {\rm P}_{R}
 \tilde{\chi}^{0 A}
 \tilde{\nu}_{X} \nonumber \\
&+ 
 \bar{d}^{\alpha} 
 \left\{
 (C^{(d)}_{R})_{\alpha}^{AX} {\rm P}_{R} 
 +
 (C^{(d)}_{L})_{\alpha}^{AX} {\rm P}_{L} 
 \right\}
 \tilde{\chi}^{-}_{A} \tilde{u}_{X}
 +
 \bar{u}^{\alpha}
 \left\{
 (C^{(u)}_{R})_{\alpha A}^{X} {\rm P}_{R}
 +
 (C^{(u)}_{L})_{\alpha A}^{X} {\rm P}_{L}
 \right\}
 \tilde{\chi}^{+ A} \tilde{d}_{X} \nonumber\\
&+
 \bar{d}^{\alpha} 
 (N^{(d)}_{R})_{\alpha A}^{X} {\rm P}_{R}
 \tilde{\chi}^{0 A}
 \tilde{d}_{X}
 +
 \bar{d}^{\alpha} 
 (N^{(d)}_{L})_{\alpha}^{AX} {\rm P}_{L}
 \tilde{\chi}^{0}_{A}
 \tilde{d}_{X}
+
\bar{u}^{\alpha} 
 (N^{(u)}_{R})_{\alpha A}^{X} {\rm P}_{R}
 \tilde{\chi}^{0 A}
 \tilde{u}_{X} 
+
\bar{u}^{\alpha} 
 (N^{(u)}_{L})_{\alpha}^{AX} {\rm P}_{L}
 \tilde{\chi}^{0}_{A}
 \tilde{u}_{X} \nonumber \\
&+\text{h.c.},
\end{align}
where the coefficients are
\begin{align}
(C^{(l)}_{R})_{\alpha}^{AX}
 =&
 -g_{2} {(U_{R}^{*})^{A}}_{1}
 {(U_{\tilde{\nu}}^{*})^{X}}_{\alpha}, \\
(C^{(l)}_{L})_{\alpha}^{AX}
 =&
 g_{2} \frac{m_{l_{\alpha}}}{\sqrt{2} m_{W} \cos\beta}
 {(U_{L}^{*})^{A}}_{2}
 {(U_{\tilde{\nu}}^{*})^{X}}_{\alpha},\\
(C^{(\nu)}_{R})_{\alpha A}^{X} 
 =&
 -g_{2} {(U_{L})_{A}}^{1}
 {(U_{\tilde{l}}^{*})^{X}}_{\alpha}, \\
(N^{(l)}_{R})_{\alpha A}^{X}
 =& 
 -\frac{g_{2}}{\sqrt{2}}
 \left[
  \left\{
   - {(U_{N})_{A}}^{2} - {(U_{N})_{A}}^{1} \tan\theta_{W}
  \right\} {(U_{\tilde{l}}^{*})^{X}}_{\alpha}
 +
 \frac{m_{l_{\alpha}}}{m_{W} \cos\beta}
 {(U_{N})_{A}}^{3}
 {(U_{\tilde{l}}^{*})^{X}}_{\alpha+3}
 \right], \\
 (N^{(l)}_{L})_{\alpha}^{AX}=&
 -\frac{g_{2}}{\sqrt{2}}
 \left\{
  \frac{m_{l_{\alpha}}}{m_{W} \cos\beta}
 {(U_{N}^{*})^{A}}_{3}
 {(U_{\tilde{l}}^{*})^{X}}_{\alpha}
 +
  2 {(U_{N}^{*})^{A}}_{1} \tan\theta_{W}
  {(U_{\tilde{l}}^{*})^{X}}_{\alpha+3}
 \right\},\\
 (N^{(\nu)}_{R})_{\alpha A}^{X} =&
 -\frac{g_{2}}{\sqrt{2}}
 \left\{
 {(U_{N})_{A}}^{2} -  {(U_{N})_{A}}^{1} \tan\theta_{W}
 \right\}
  {(U_{\tilde{\nu}}^{*})^{X}}_{\alpha}, \\
(C^{(d)}_{R})_{\alpha}^{AX} =& 
 g_{2} 
 \left\{
  - {( U_{R}^{*})^{A}}_{1}
    {(U_{\tilde{u}}^{*})^{X}}_{\alpha}
  +
  \frac{m_{u_{\alpha}}}{\sqrt{2} m_{W} \sin \beta}
  {( U_{R}^{*})^{A}}_{2}
  {(U_{\tilde{u}}^{*})^{X}}_{\alpha+3}
 \right\},\\
(C^{(d)}_{L})_{\alpha}^{AX} =& 
 g_{2} 
 \frac{m_{d_{\alpha}}}{\sqrt{2} m_{W} \cos \beta}
 {( U_{L}^{*})^{A}}_{2}
 {(U_{\tilde{u}}^{*})^{X}}_{\alpha}
,\\
(C^{(u)}_{R})_{\alpha A}^{X} =& 
 g_{2} 
 \left\{
  -  {( U_{L})_{A}}^{1}
    {(U_{\tilde{d}}^{*})^{X}}_{\alpha}
 +
 \frac{m_{d_{\alpha}}}{\sqrt{2} m_{W} \cos \beta}
  {( U_{L})_{A}}^{2}
 {(U_{\tilde{d}}^{*})^{X}}_{\alpha+3}
 \right\}
,\\
(C^{(u)}_{L})_{\alpha A}^{X} =& 
 g_{2} 
 \frac{m_{u_{\alpha}}}{\sqrt{2} m_{W} \sin \beta}
 {( U_{R})_{A}}^{2}
 {(U_{\tilde{d}}^{*})^{X}}_{\alpha}
 ,\\
(N^{(d)}_{R})_{\alpha A}^{X} =& 
 -\frac{g_{2}}{\sqrt{2}}
 \left[
  \left\{
   - {(U_{N})_{A}}^{2} + \frac{1}{3} {(U_{N})_{A}}^{1} \tan\theta_{W}
  \right\} {(U_{\tilde{d}}^{*})^{X}}_{\alpha}
 +
 \frac{m_{d_{\alpha}}}{m_{W} \cos\beta}
 {(U_{N})_{A}}^{3}
 {(U_{\tilde{d}}^{*})^{X}}_{\alpha+3}
 \right] 
,\\
(N^{(d)}_{L})_{\alpha}^{AX} =& 
 -\frac{g_{2}}{\sqrt{2}}
 \left\{
  \frac{m_{d_{\alpha}}}{m_{W} \cos\beta}
 {(U_{N}^{*})^{A}}_{3}
 {(U_{\tilde{d}}^{*})^{X}}_{\alpha}
 +
  \frac{2}{3} {(U_{N}^{*})^{A}}_{1} \tan\theta_{W}
  {(U_{\tilde{d}}^{*})^{X}}_{\alpha+3}
 \right\}
,\\
(N^{(u)}_{R})_{\alpha A}^{X} =& 
 -\frac{g_{2}}{\sqrt{2}}
 \left[
 \left\{
 {(U_{N})_{A}}^{2} + \frac{1}{3}{(U_{N})_{A}}^{1} \tan\theta_{W}
 \right\}
  {(U_{\tilde{u}}^{*})^{X}}_{\alpha}
 +
 \frac{m_{u_{\alpha}}}{m_{W} \sin\beta}
 {(U_{N})_{A}}^{4}
 {(U_{\tilde{u}}^{*})^{X}}_{\alpha+3}
 \right], \\
(N^{(u)}_{L})_{\alpha}^{AX} =&
 -\frac{g_{2}}{\sqrt{2}}
 \left\{
  \frac{m_{u_{\alpha}}}{m_{W} \sin\beta}
 {(U_{N}^{*})^{A}}_{4}
 {(U_{\tilde{u}}^{*})^{X}}_{\alpha}
 -
 \frac{4}{3}
 {(U_{N}^{*})^{A}}_{1} \tan\theta_{W}
  {(U_{\tilde{u}}^{*})^{X}}_{\alpha+3}
 \right\}.
\end{align}

\section{Details for $\epsilon$'s}
\label{app:calculation}
We here show the explicit form of the nLFV parameters
$\epsilon^{s}_{\mu\tau}$, $\epsilon^{m}_{\mu\tau}$ and
$\epsilon^{d}_{\mu\tau}$ in the MSSMRN.
They are calculated from 1-loop diagrams.

\subsection{For $\epsilon^{s}_{\mu\tau}$}
\begin{figure}[tb]
\includegraphics[width=16cm]{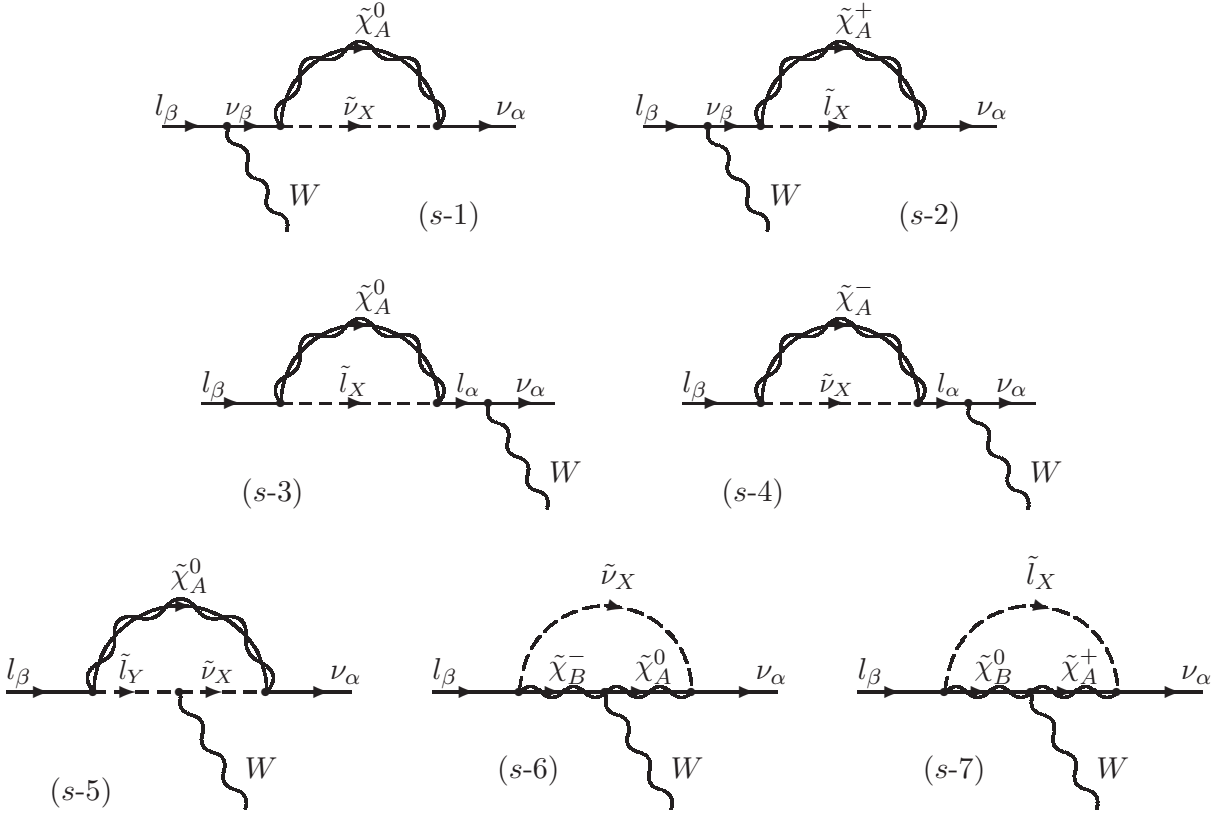}
\caption{Diagrams which contribute to $(\epsilon^{s}_{\mu\tau})_{\text{$W$-penguin}}$.}
\label{Fig:penguin-epsilonS}
\end{figure}

The effective coupling comprises two kinds of contribution; the one
comes from the penguin-type diagram associated with $W$ boson and 
the other is the box-type diagram:
\begin{align}
\epsilon^{s}_{\mu\tau} = 
 (\epsilon^{s}_{\mu\tau})_{\text{$W$-penguin}}
 +
 (\epsilon^{s}_{\mu\tau})_{\text{box}}.
\end{align}
The penguin-part is represented as 
\begin{align}
(\epsilon^{s}_{\mu\tau})_{\text{$W$-penguin}} 
= \sum_{i} \mathcal{A}^{(s\text{-}i)}_{\mu\tau},
\end{align}
and each contribution which is shown in Fig. \ref{Fig:penguin-epsilonS}
is calculated to be
\begin{align}
\mathcal{A}^{\text{($s$-1)}}_{\beta \alpha}
=&
 \frac{1}{(4\pi)^{2}}
 (N^{(\nu)}_{R})_{\alpha A}^{X}
 (N^{(\nu)*}_{R})^{\beta A}_{X}
D(m_{\tilde{\nu}_{X}}, M_{\tilde{\chi}^{0}_{A}}),
\label{eq:source-W-penguin-Diagram-i-c}\\
\mathcal{A}^{\text{($s$-2)}}_{\beta \alpha}
=&
 \frac{1}{(4\pi)^{2}}
  (C^{(\nu)}_{R})_{\alpha A}^{X}
  (C^{(\nu)*}_{R})^{\beta A}_{X} 
D(m_{\tilde{l}_{X}}, M_{\tilde{\chi}^{+}_{A}}),
\label{eq:source-W-penguin-Diagram-ii-c}\\
\mathcal{A}^{\text{($s$-3)}}_{\beta \alpha}=&
 \frac{1}{(4\pi)^{2}}
  \frac{m_{l_{\beta}}^{2}}
       {m_{l_{\beta}}^{2} - m_{l_{\alpha}}^{2}}
 (N^{(l)}_{R})_{\alpha A}^{X} 
 (N^{(l)*}_{R})^{\beta A}_{X} 
D(m_{\tilde{l}_{X}}, M_{\tilde{\chi}^{0}_{A}}),
\label{eq:source-W-penguin-Diagram-i-b} \\
\mathcal{A}^{\text{($s$-4)}}_{\beta \alpha}
=&
 \frac{1}{(4\pi)^{2}}
  \frac{m_{l_{\beta}}^{2}}
       {m_{l_{\beta}}^{2} - m_{l_{\alpha}}^{2}}
 (C^{(l)}_{R})_{\alpha}^{AX}
 (C^{(l)*}_{R})^{\beta}_{AX} 
D(m_{\tilde{\nu}_{X}}, M_{\tilde{\chi}^{-}_{A}}),
\label{eq:source-W-penguin-Diagram-ii-b} \\
\mathcal{A}^{\text{($s$-5)}}_{\beta \alpha} =&
-
 \frac{1}{(4\pi)^{2}}
(N^{(\nu)}_{R})_{\alpha A}^{X}
 (N^{(l)*}_{R})^{\beta A}_{Y}
 {(U_{\tilde{\nu}})_{X}}^{\gamma = \text{1-3}}  {(U_{\tilde{l}}^{{\rm
 \dagger}})_{\gamma=\text{1-3}}}^{Y}
E(m_{\tilde{\nu}_{X}}, m_{\tilde{l}_{Y}}, M_{\tilde{\chi}^{0}_{A}}),
\label{eq:source-W-penguin-Diagram-i-a} \\
\mathcal{A}^{\text{($s$-6)}}_{\beta \alpha}
=&
\sqrt{2}
 \frac{1}{(4\pi)^{2}}
 (N^{(\nu)}_{R})_{\alpha A}^{X} 
 (C^{(l)*}_{R})^{\beta}_{BX}  \nonumber \\
& \times
 \left[
 \delta^{AA} {(O^{R*})_{A}}^{B} 
 G 
 \left({m_{\tilde{\nu}_{X}},
 M_{\tilde{\chi}^{0}_{A}}, M_{\tilde{\chi}^{-}_{B}}}
 \right) 
 -
 {(O^{L*})}^{AB} \frac{1}{2} 
 F
 \left({m_{\tilde{\nu}_{X}},
 M_{\tilde{\chi}^{0}_{A}}, M_{\tilde{\chi}^{-}_{B}}}
 \right) 
 \right],
\label{eq:source-W-penguin-Diagram-iii} \\
\mathcal{A}^{\text{($s$-7)}}_{\beta\alpha}
=&
-
\sqrt{2}
\frac{1}{(4\pi)^{2}}
 (C^{(\nu)}_{R})_{\alpha A}^{X} 
 (N^{(l)*}_{R})^{\beta B}_{X}  \nonumber \\
&\times 
\left[
 \delta_{BB}
 {(O^{L*})^{BA}}  
 G
 \left({m_{\tilde{l}_{X}},
 M_{\tilde{\chi}^{+}_{A}}, M_{\tilde{\chi}^{0}_{B}}}
 \right) 
 -
 {(O^{R*})_{B}}^{A} \frac{1}{2} 
 F
 \left({m_{\tilde{l}_{X}},
 M_{\tilde{\chi}^{+}_{A}}, M_{\tilde{\chi}^{0}_{B}}}
 \right) 
\right],
\label{eq:source-W-penguin-Diagram-iv}
\end{align}
where the functions $D$, $E$, $F$ and $G$ are defined as
\begin{align}
&D(m_{X}, M_{A}) \equiv
 \frac{1}{4} \frac{1}{(1-x_{AX})^{2}}
 \left\{ 1
 -4 x_{AX} 
 + 3 x_{AX}^{2} 
 - 2 x_{AX}^{2} 
 \ln x_{AX}
 - 2 (1-x_{AX})^{2} \ln m_{X}^{2}
 \right\}, \\
&E(m_{X}, m_{Y}, M_{A}) \equiv
 \frac{1}{2}
 \frac{1}{x_{AY} - x_{AX}}
 \left(
  \frac{ x_{AY} \ln  x_{AX}}{1 - x_{AX} }
 -
  \frac{ x_{AX} \ln  x_{AY}}{1 - x_{AY} }
 \right), \\
&F(m_{X}, M_{A}, M_{B}) \equiv
 \ln x_{AX} +
 \frac{1}{x_{AX} - x_{BX}}
 \left(
 \frac{x_{AX}^{2} \ln x_{AX}}{1-x_{AX}}
 -
 \frac{x_{BX}^{2} \ln x_{BX}}{1-x_{BX}}
 \right), \\
&G(m_{X}, M_{A}, M_{B})  \equiv
 \sqrt{x_{AX} x_{BX}}
 \frac{1}{x_{AX} - x_{BX}}
 \left(
 \frac{x_{AX} \ln x_{AX}}{1-x_{AX}}
 -
 \frac{x_{BX} \ln x_{BX}}{1-x_{BX}}
 \right),
\end{align}
with $x_{AX} \equiv M_{A}^{2} / m_{X}^{2}$.
The couplings for the chargino-neutralino-$W$-boson interaction, $O^{L}$ and $O^{R}$, 
are given as\cite{HaberKane}
\begin{align}
(O^{L})_{AB} &= 
  - \frac{1}{\sqrt{2}} {(U_{N})_{A}}^{4} {(U_{R})_{B}}^{2}
                     + {(U_{N})_{A}}^{2} {(U_{R})_{B}}^{1}, \\
{(O^{R})^{A}}_{B} &= 
 \frac{1}{\sqrt{2}} {(U_{N}^{*})^{A}}_{3} {(U_{L})_{B}}^{2}
                  + {(U_{N}^{*})^{A}}_{2} {(U_{L})_{B}}^{1}.
\end{align}  

\begin{figure}[thb]
\includegraphics[width=12cm]{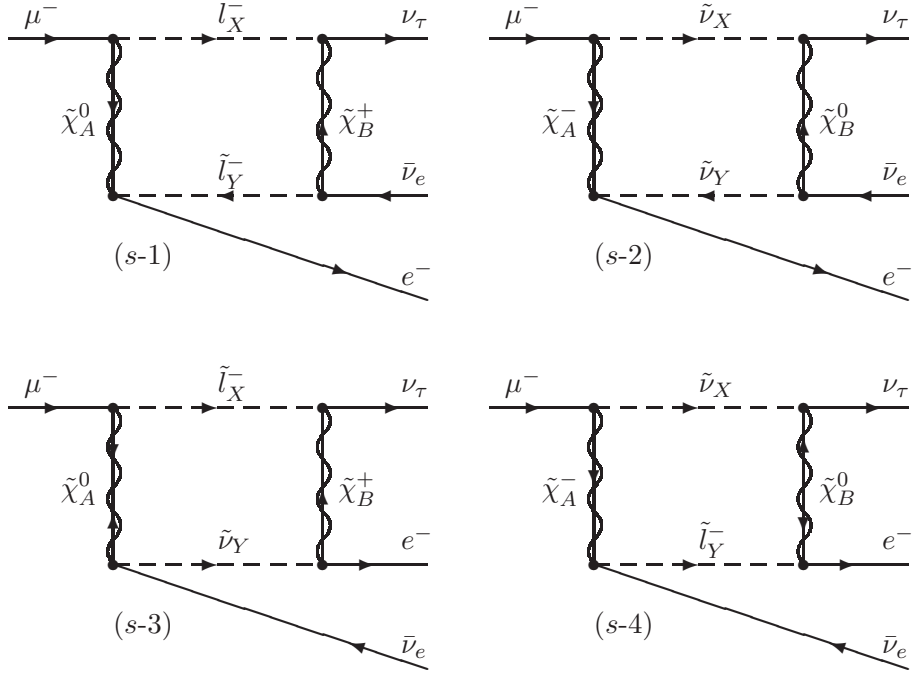}
\caption{Diagrams which contribute to $(\epsilon^{s}_{\mu\tau})_{\text{box}}$.}
\label{Fig:box-epsilonS}
\end{figure}
The box-part is represented as
\begin{align}
(\epsilon^{s}_{\mu\tau})_{\text{box}} = \sum_{i} \mathcal{B}^{(s\text{-}i)}_{\mu\tau},
\end{align}
and each contribution, which is shown in Fig. \ref{Fig:box-epsilonS}, is
calculated to be 
\begin{align}
\mathcal{B}^{(s\text{-}1)}_{\beta\alpha} 
=& 
  \frac{1}{8\sqrt{2} G_{F}}
 J_{4}
 \left(
 M_{\tilde{\chi}^{0}_{A}},
 M_{\tilde{\chi}^{+}_{B}},
 m_{\tilde{l}_{X}},
 m_{\tilde{l}_{Y}}
 \right)
 (N_{R}^{(l) *})_{X}^{\beta A}
 (N_{R}^{(l)})_{eA}^{Y} 
 (C_{R}^{(\nu) *})_{Y}^{eB}
 (C_{R}^{(\nu)})_{\alpha B}^{X} 
 ,\label{eq:box-epsilonS-i} \\
\mathcal{B}^{(s\text{-}2)}_{\beta\alpha}
=& 
 \frac{1}{8\sqrt{2} G_{F}}
 J_{4}
 \left(
 M_{\tilde{\chi}^{-}_{A}},
 M_{\tilde{\chi}^{0}_{B}},
 m_{\tilde{\nu}_{X}},
 m_{\tilde{\nu}_{Y}}
 \right)
 (C_{R}^{(l) *})^{\beta}_{AX}
 (C_{R}^{(l)})_{e}^{AY} 
 (N_{R}^{(\nu) *})_{Y}^{eB}
 (N_{R}^{(\nu)})_{\alpha B}^{X}
, \label{eq:box-epsilonS-ii} \\
\mathcal{B}^{(s\text{-}3)}_{\beta\alpha}
=&
\frac{1}{4\sqrt{2} G_{F}}
 I_{4}
 \left(
 M_{\tilde{\chi}^{0}_{A}},
 M_{\tilde{\chi}^{+}_{B}},
 m_{\tilde{l}_{X}},
 m_{\tilde{\nu}_{Y}}
 \right)
 M_{\tilde{\chi}^{0}_{A}} M_{\tilde{\chi}^{+}_{B}}
 (N_{R}^{(l) *})_{X}^{\beta A}
 (N_{R}^{(\nu)*})_{Y}^{eA}
 (C_{R}^{(\nu)})^{X}_{\alpha B}
 (C_{R}^{(l)})_{e}^{BY}  
,\label{eq:box-epsilonS-iii} \\
\mathcal{B}^{(s\text{-}4)}_{\beta\alpha}
=&
\frac{1}{4\sqrt{2} G_{F}}
 I_{4}  
 \left(
 M_{\tilde{\chi}^{-}_{A}},
 M_{\tilde{\chi}^{0}_{B}},
 m_{\tilde{\nu}_{X}},
 m_{\tilde{l}_{Y}}
 \right)
 M_{\tilde{\chi}^{-}_{A}} M_{\tilde{\chi}^{0}_{B}}
 ( C_{R}^{(l) *} )^{\beta}_{AX}
  (C_{R}^{(\nu)*})_{X}^{eA}
  (N_{R}^{(\nu)})_{\alpha B}^{X}
  (N_{R}^{(l)})_{eB}^{Y}  
, \label{eq:box-epsilonS-iv}
\end{align}
where $I_{4}$ and $J_{4}$ are the functions which are given as
\begin{align}
I_{4}(M_{A}, M_{B}, m_{X}, m_{Y}) 
& \equiv \int \frac{{\rm d}^{4} k }{(2 \pi)^{4} {\rm i}}
 \frac{1}{(k^{2} - M_{A}^{2})(k^{2} - M_{B}^{2})(k^{2} -
 m_{X}^{2})(k^{2} - m_{Y}^{2})},\\
J_{4}(M_{A}, M_{B}, m_{X}, m_{Y}) 
& \equiv \int \frac{{\rm d}^{4} k }{(2 \pi)^{4} {\rm i}}
 \frac{k^{2}}{(k^{2} - M_{A}^{2})(k^{2} - M_{B}^{2})(k^{2} -
 m_{X}^{2})(k^{2} - m_{Y}^{2})}.
\end{align}

\subsection{For $\epsilon^{m}_{\mu\tau}$}
Since the matter of the Earth is the neutral for $U(1)_{\text{em}}$,
there is no contribution to $\epsilon^{m}_{\mu\tau}$ from
photon-penguin diagrams.
The $Z$-penguin contribution associated with a proton and that with
a neutron cancel each other out. 
The contributions which need to be taken into account
are 
the $Z$-penguin contribution associated with an electron and 
the box contributions:
\begin{gather}
\epsilon^{m,e}_{\mu\tau} =
(\epsilon^{m,e}_{\mu\tau})_{\text{$Z$-penguin}}
 +
(\epsilon^{m,e}_{\mu\tau})_{\text{box}}, \\
\epsilon^{m,u}_{\mu\tau} =
(\epsilon^{m,u}_{\mu\tau})_{\text{box}}, \\
\epsilon^{m,d}_{\mu\tau} =
(\epsilon^{m,d}_{\mu\tau})_{\text{box}}.
\end{gather}

\begin{figure}[htb]
\includegraphics[width=16cm]{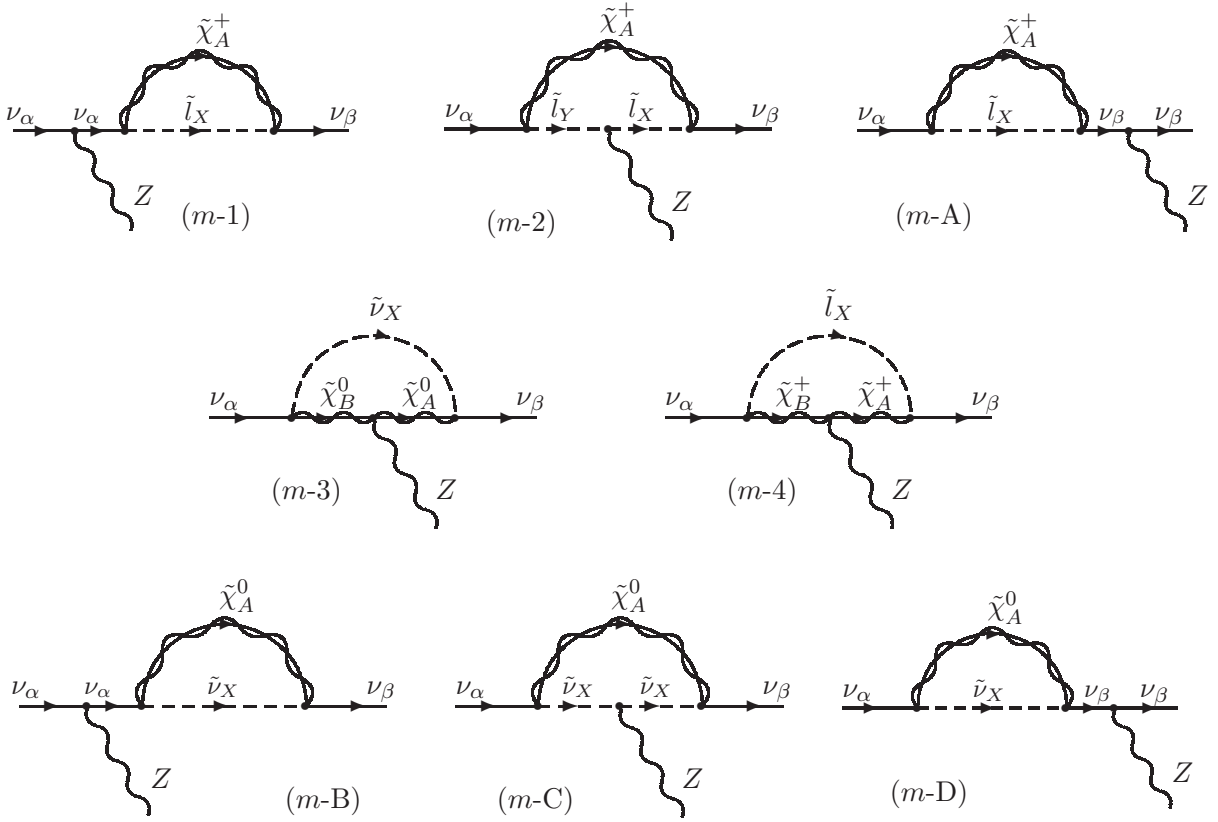}
\caption{Diagrams which contributes to
 $(\epsilon^{m,e}_{\mu\tau})_{\text{$Z$-penguin}}$.
 The diagrams ($m$-A) and ($m$-D) are not counted into
 $\epsilon^{m}_{\mu\tau}$, which are included in
 $\epsilon^{s}_{\mu\tau}$ (cf. ($s$-2) and ($s$-1) in Fig. \ref{Fig:penguin-epsilonS}).
 The diagrams ($m$-B) and ($m$-C) cancel each other out
 after the off-shell prescription (Eq. \eqref{eq:masslessNeutrino}).}
\label{Fig:penguin-epsilonM}
\end{figure}

The penguin contribution consists of diagrams which are drawn 
in Fig. \ref{Fig:penguin-epsilonM}, 
\begin{align}
(\epsilon^{m,e}_{\mu\tau})_{\text{$Z$-penguin}}
= \sum_{i} \mathcal{A}^{(m\text{-}i)}_{\mu\tau}.
\end{align}
Each diagrams is calculated to be
\begin{align}
\mathcal{A}^{(m\text{-}1)}_{\alpha\beta}
=&
 \frac{1}{(4 \pi)^{2}}
 \frac{1}{2}
 \left( 2 \sin^{2} \theta_{W} -1 \right)
 ( C^{(\nu)}_{R} )^{X}_{\beta A}
 ( C^{(\nu)*}_{R})_{X}^{\alpha A} 
 D(m_{\tilde{l}_{X}}, M_{\tilde{\chi}^{+}_{A}})
\label{eq:Z-penguin-epsilonM-iii-b+iii-c},\\
\mathcal{A}^{(m\text{-}2)}_{\alpha\beta}
=& -
 \frac{1}{(4 \pi)^{2}}
 \left( 2 \sin^{2} \theta_{W} -1 \right)
 (C^{(\nu)}_{R})^{X}_{\beta A}
 (C^{(\nu)*}_{R})^{\alpha A}_{Y} 
 \nonumber \\
& \quad \times
 \left\{
 \left(
   -\frac{1}{2} + \sin^{2} \theta_{W}
 \right)
 {(U_{\tilde{l}})_{X}}^{\gamma=\text{1-3}}
 {(U_{\tilde{l}}^{\dagger})_{\gamma=\text{1-3}}}^{Y}
 +
 \left(
 \sin^{2} \theta_{W}
 \right)
 {(U_{\tilde{l}})_{X}}^{\gamma=\text{4-6}}
 {(U_{\tilde{l}}^{\dagger})_{\gamma=\text{4-6}}}^{Y}
 \right\} \nonumber \\
&  \qquad \times
E(m_{\tilde{l}_{X}}, m_{\tilde{l}_{Y}}, M_{\tilde{\chi}^{+}_{A}})
\label{eq:Z-penguin-epsilonM-iii-a} \\
\mathcal{A}^{(m\text{-}3)}_{\alpha\beta}
=&
 - \frac{1}{(4\pi)^{2}} 
 \left( 2 \sin^{2} \theta_{W} -1 \right)
 (N^{(\nu)}_{R})^{X}_{\beta A}
 (N^{(\nu)*}_{R})^{\alpha B}_{X} \nonumber \\
 &\quad \times 
 \left[
 {(O''^{L})_{A}}^{B}  
 G
 \left(
 m_{\tilde{\nu}_{X}}, M_{\tilde{\chi}^{0}_{A}}, M_{\tilde{\chi}^{0}_{B}}
 \right)
  -
  \frac{1}{2} {(O''^{R})^{A}}_{B}
  F
 \left(
 m_{\tilde{\nu}_{X}}, M_{\tilde{\chi}^{0}_{A}}, M_{\tilde{\chi}^{0}_{B}}
 \right)
 \right],
\label{eq:Z-penguin-epsilonM-ii}\\
\mathcal{A}^{(m\text{-}4)}_{\alpha\beta}
=&
  - \frac{1}{(4 \pi)^{2}}
 \left( 2 \sin^{2} \theta_{W} -1 \right)
 (C^{(\nu)}_{R})^{X}_{\beta A} 
 (C^{(\nu)*}_{R})^{\alpha B}_{X} \nonumber \\
&\quad \times
 \biggl[
  {(O'^{L})^{A}}_{B} 
 G
 \left(
 m_{\tilde{l}_{X}}, M_{\tilde{\chi}^{+}_{A}}, M_{\tilde{\chi}^{+}_{B}}
 \right)
 -
  \frac{1}{2} {(O'^{R})^{A}}_{B} 
 F
 \left(
 m_{\tilde{l}_{X}}, M_{\tilde{\chi}^{+}_{A}}, M_{\tilde{\chi}^{+}_{B}}
 \right)
 \biggr].
\label{eq:Z-penguin-epsilonM-iv}
\end{align}
The couplings for the chargino-chargino-$Z$-boson 
and neutralino-neutralino-$Z$-boson are\cite{HaberKane}
\begin{align}
{(O'^{L})^{A}}_{B} =& 
 - {(U_{R}^{*})^{A}}_{1} {(U_{R})_{B}}^{1}
 - \frac{1}{2} {(U_{R}^{*})^{A}}_{2} {(U_{R})_{B}}^{2} 
 + \delta^{A}_{B} \sin^{2} \theta_{W},  \\
{(O'^{R})^{A}}_{B} =& 
 - {(U_{L}^{*})^{A}}_{1} {(U_{L})_{B}}^{1}
 - \frac{1}{2} {(U_{L}^{*})^{A}}_{2} {(U_{L})_{B}}^{2} 
 + \delta^{A}_{B} \sin^{2} \theta_{W}, \\
{(O''^{L})_{A}}^{B} =& 
 - \frac{1}{2} {(U_{N})_{A}}^{3} {(U_{N}^{*})^{B}}_{3}
 + \frac{1}{2} {(U_{N})_{A}}^{4} {(U_{N}^{*})^{B}}_{4}, \\
{(O''^{R})^{A}}_{B} =& - {(O''^{L *})^{A}}_{B}.   
\end{align}
Here, we take into account the procedure to resolve the double counting
problem
which is explained in Sec. \ref{sec:analytic-calc}.

The box contribution associated with the electron in the Earth's matter is
\begin{align}
(\epsilon^{m,e}_{\mu\tau})_{\text{box}} = \sum_{i} \mathcal{B}^{(m,e\text{-}i)}_{\mu\tau},
\end{align}
where
\begin{align}
\mathcal{B}^{(m,e\text{-}1)}_{\alpha\beta}
 =&
 \frac{1}{8\sqrt{2} G_{F}}
 J_{4}
 \left(
 M_{\tilde{\chi}^{0}_{A}}, M_{\tilde{\chi}^{0}_{B}},
 m_{\tilde{\nu}_{X}}, m_{\tilde{l}_{Y}}
 \right)
 (N^{(\nu)}_{R})^{X}_{\beta B}
 (N^{(l)*}_{R})^{eB}_{Y}
 (N^{(l)}_{R})^{Y}_{eA}
 (N^{(\nu)*}_{R})^{\alpha A}_{X}
,\\
\mathcal{B}^{(m,e\text{-}2)}_{\alpha\beta}
=&
 -
 \frac{1}{4\sqrt{2} G_{F}}
  I_{4}
 \left(
 M_{\tilde{\chi}^{0}_{A}}, M_{\tilde{\chi}^{0}_{B}},
 m_{\tilde{\nu}_{X}}, m_{\tilde{l}_{Y}}
 \right)
 M_{\tilde{\chi}^{0}_{A}}
 M_{\tilde{\chi}^{0}_{B}}
 (N^{(\nu)}_{R})^{X}_{\beta B}
 (N^{(l)*}_{L})^{e}_{BY}
 (N^{(l)}_{L})^{AY}_{e}
 (N^{(\nu)*}_{R})^{\alpha A}_{X}
 ,\\ 
\mathcal{B}^{(m,e\text{-}3)}_{\alpha\beta}
=&
 \frac{1}{4\sqrt{2} G_{F}}
  I_{4}
 \left(
 M_{\tilde{\chi}^{0}_{A}}, M_{\tilde{\chi}^{0}_{B}},
 m_{\tilde{\nu}_{X}}, m_{\tilde{l}_{Y}}
 \right)
 M_{\tilde{\chi}^{0}_{A}}
 M_{\tilde{\chi}^{0}_{B}}
 (N^{(\nu)}_{R})^{X}_{\beta B}
 (N^{(l)}_{R})^{Y}_{eB} 
 (N^{(\nu)*}_{R})^{\alpha A}_{X}
 (N^{(l)*}_{R})^{eA}_{Y}
,\\
\mathcal{B}^{(m,e\text{-}4)}_{\alpha\beta}
 =&
  -
 \frac{1}{8\sqrt{2} G_{F}}
 J_{4}
 \left(
 M_{\tilde{\chi}^{0}_{A}}, M_{\tilde{\chi}^{0}_{B}},
 m_{\tilde{\nu}_{X}}, m_{\tilde{l}_{Y}}
 \right)
 (N^{(\nu)}_{R})^{X}_{\beta B}
 (N^{(l)}_{R})^{Y}_{eB} 
 (N^{(\nu)*}_{R})^{\alpha A}_{X}
 (N^{(l)*}_{R})^{eA}_{Y}
,\\
\mathcal{B}^{(m,e\text{-}5)}_{\alpha\beta}
 =&
 \frac{1}{4\sqrt{2} G_{F}}
 I_{4}
 \left(
 M_{\tilde{\chi}^{+}_{A}}, M_{\tilde{\chi}^{+}_{B}},
 m_{\tilde{l}_{X}}, m_{\tilde{\nu}_{Y}}
 \right)
 M_{\tilde{\chi}^{+}_{A}}
 M_{\tilde{\chi}^{+}_{B}}
 (C^{(\nu)}_{R})^{X}_{\beta B}
 (C^{(l)}_{R})^{BY}_{e}
 (C^{(l)*}_{R})^{e}_{AY}
 (C^{(\nu)*}_{R})^{\alpha A}_{X} 
,\\
\mathcal{B}^{(m,e\text{-}6)}_{\alpha\beta}
 =&
 -\frac{1}{8\sqrt{2} G_{F}}
 J_{4}
 \left(
 M_{\tilde{\chi}^{+}_{A}}, M_{\tilde{\chi}^{+}_{B}},
 m_{\tilde{l}_{X}}, m_{\tilde{\nu}_{Y}}
 \right)
 (C^{(\nu)}_{R})^{X}_{\beta B}
 (C^{(l)}_{L})^{BY}_{e}
 (C^{(l)*}_{L})^{e}_{AY}
 (C^{(\nu)*}_{R})^{\alpha A}_{X} 
,\\
\mathcal{B}^{(m,e\text{-}7)}_{\alpha\beta}
 =&
 \frac{1}{4\sqrt{2} G_{F}}
 I_{4}
 \left(
 M_{\tilde{\chi}^{+}_{A}}, M_{\tilde{\chi}^{0}_{B}},
 m_{\tilde{l}_{X}}, m_{\tilde{\nu}_{Y}}
 \right)
 M_{\tilde{\chi}^{+}_{A}}
 M_{\tilde{\chi}^{0}_{B}}
 (N^{(l)}_{R})^{X}_{eB}
 (N^{(\nu)}_{R})^{Y}_{\beta B}
 (C^{(l)*}_{R})^{e}_{AY}
 (C^{(\nu)*}_{R})^{\alpha A}_{X}
,\\
\mathcal{B}^{(m,e\text{-}8)}_{\alpha\beta}
 =&
  \frac{1}{8\sqrt{2} G_{F}}
 J_{4}
 \left(
 M_{\tilde{\chi}^{+}_{A}}, M_{\tilde{\chi}^{0}_{B}},
 m_{\tilde{l}_{X}}, m_{\tilde{\nu}_{Y}}
 \right)
 (N^{(l)}_{L})^{BX}_{e}
 (N^{(\nu)}_{R})^{Y}_{\beta B}
 (C^{(l)*}_{L})^{e}_{AY}
 (C^{(\nu)*}_{R})^{\alpha A}_{X}
,\\
\mathcal{B}^{(m,e\text{-}9)}_{\alpha\beta}
 =&
 \frac{1}{4\sqrt{2} G_{F}}
 I_{4}
 \left(
 M_{\tilde{\chi}^{0}_{A}}, M_{\tilde{\chi}^{-}_{B}},
 m_{\tilde{\nu}_{X}}, m_{\tilde{l}_{Y}}
 \right)
 M_{\tilde{\chi}^{0}_{A}}
 M_{\tilde{\chi}^{-}_{B}}
 (C^{(l)}_{R})^{BX}_{e}
 (C^{(\nu)}_{R})^{Y}_{\beta B}
 (N^{(l)*}_{R})^{eA}_{Y}
 (N^{(\nu)*}_{R})^{\alpha A}_{X}
,\\
\mathcal{B}^{(m,e\text{-}10)}_{\alpha\beta}
=&
 \frac{1}{8\sqrt{2} G_{F}}
 J_{4}
 \left(
 M_{\tilde{\chi}^{0}_{A}}, M_{\tilde{\chi}^{-}_{B}},
 m_{\tilde{\nu}_{X}}, m_{\tilde{l}_{Y}}
 \right)
 (C^{(l)}_{L})^{BX}_{e}
 (C^{(\nu)}_{R})^{Y}_{\beta B} 
 (N^{(l)*}_{L})^{eA}_{Y}
 (N^{(\nu)*}_{R})^{\alpha A}_{X}
.
\end{align}

The box contribution associated with the down-quark in the matter of the
Earth is
\begin{align}
(\epsilon^{m,d}_{\mu\tau})_{\text{box}} = \sum_{i} \mathcal{B}^{(m,d\text{-}i)}_{\mu\tau},
\end{align}
where
\begin{align}
\mathcal{B}^{(m,d\text{-}1)}_{\alpha\beta}
 =&
 \frac{1}{8\sqrt{2} G_{F}}
 J_{4}
 \left(
 M_{\tilde{\chi}^{0}_{A}}, M_{\tilde{\chi}^{0}_{B}},
 m_{\tilde{\nu}_{X}}, m_{\tilde{d}_{Y}}
 \right)
 (N^{(\nu)}_{R})^{X}_{\beta B}
 (N^{(d)*}_{R})^{dB}_{Y}
 (N^{(d)}_{R})^{Y}_{dA}
 (N^{(\nu)*}_{R})^{\alpha A}_{X}
,\\
\mathcal{B}^{(m,d\text{-}2)}_{\alpha\beta}
=&
 -
 \frac{1}{4\sqrt{2} G_{F}}
 I_{4}
 \left(
 M_{\tilde{\chi}^{0}_{A}}, M_{\tilde{\chi}^{0}_{B}},
 m_{\tilde{\nu}_{X}}, m_{\tilde{d}_{Y}}
 \right)
 M_{\tilde{\chi}^{0}_{A}}
 M_{\tilde{\chi}^{0}_{B}}
 (N^{(\nu)}_{R})^{X}_{\beta B}
 (N^{(d)*}_{L})^{d}_{BY}
 (N^{(d)}_{L})^{AY}_{d}
 (N^{(\nu)*}_{R})^{\alpha A}_{X}
,\\
\mathcal{B}^{(m,d\text{-}3)}_{\alpha\beta}
=&
 \frac{1}{4\sqrt{2} G_{F}}
  I_{4}
 \left(
 M_{\tilde{\chi}^{0}_{A}}, M_{\tilde{\chi}^{0}_{B}},
 m_{\tilde{\nu}_{X}}, m_{\tilde{d}_{Y}}
 \right)
 M_{\tilde{\chi}^{0}_{A}}
 M_{\tilde{\chi}^{0}_{B}}
 (N^{(\nu)}_{R})^{X}_{\beta B}
 (N^{(d)}_{R})^{Y}_{dB} 
 (N^{(\nu)*}_{R})^{\alpha A}_{X}
 (N^{(d)*}_{R})^{dA}_{Y}
,\\
\mathcal{B}^{(m,d\text{-}4)}_{\alpha\beta}
 =&
  -
 \frac{1}{8\sqrt{2} G_{F}}
  J_{4}
 \left(
 M_{\tilde{\chi}^{0}_{A}}, M_{\tilde{\chi}^{0}_{B}},
 m_{\tilde{\nu}_{X}}, m_{\tilde{d}_{Y}}
 \right)
 (N^{(\nu)}_{R})^{X}_{\beta B}
 (N^{(d)}_{R})^{Y}_{dB} 
 (N^{(\nu)*}_{R})^{\alpha A}_{X}
 (N^{(d)*}_{R})^{dA}_{Y}
,\\
\mathcal{B}^{(m,d\text{-}5)}_{\alpha\beta}
 =&
 \frac{1}{4\sqrt{2} G_{F}}
 I_{4}
 \left(
 M_{\tilde{\chi}^{+}_{A}}, M_{\tilde{\chi}^{+}_{B}},
 m_{\tilde{l}_{X}}, m_{\tilde{u}_{Y}}
 \right)
 M_{\tilde{\chi}^{+}_{A}}
 M_{\tilde{\chi}^{+}_{B}}
 (C^{(\nu)}_{R})^{X}_{\beta B}
 (C^{(d)}_{R})^{BY}_{d}
 (C^{(d)*}_{R})^{d}_{AY}
 (C^{(\nu)*}_{R})^{\alpha A}_{X} 
,\\
\mathcal{B}^{(m,d\text{-}6)}_{\alpha\beta}
 =&
 -\frac{1}{8\sqrt{2} G_{F}}
 J_{4}
 \left(
 M_{\tilde{\chi}^{+}_{A}}, M_{\tilde{\chi}^{+}_{B}},
 m_{\tilde{l}_{X}}, m_{\tilde{u}_{Y}}
 \right)
 (C^{(\nu)}_{R})^{X}_{\beta B}
 (C^{(d)}_{L})^{BY}_{d}
 (C^{(d)*}_{L})^{d}_{AY}
 (C^{(\nu)*}_{R})^{\alpha A}_{X} 
.
\end{align}

The box contribution associated with the up-quark in the matter of the Earth
is
\begin{align}
(\epsilon^{m,u}_{\mu\tau})_{\text{box}} = \sum_{i} \mathcal{B}^{(m,u\text{-}i)}_{\mu\tau},
\end{align}
where
\begin{align}
\mathcal{B}^{(m,u\text{-}1)}_{\alpha\beta}
 =&
 \frac{1}{8\sqrt{2} G_{F}} 
 J_{4}
 \left(
 M_{\tilde{\chi}^{0}_{A}}, M_{\tilde{\chi}^{0}_{B}},
 m_{\tilde{\nu}_{X}}, m_{\tilde{u}_{Y}}
 \right)
 (N^{(\nu)}_{R})^{X}_{\beta B}
 (N^{(u)*}_{R})^{uB}_{Y}
 (N^{(u)}_{R})^{Y}_{uA}
 (N^{(\nu)*}_{R})^{\alpha A}_{X}
,\\
\mathcal{B}^{(m,u\text{-}2)}_{\alpha\beta}
=&
 -
 \frac{1}{4\sqrt{2} G_{F}}
 I_{4}
 \left(
 M_{\tilde{\chi}^{0}_{A}}, M_{\tilde{\chi}^{0}_{B}},
 m_{\tilde{\nu}_{X}}, m_{\tilde{u}_{Y}}
 \right)
 M_{\tilde{\chi}^{0}_{A}}
 M_{\tilde{\chi}^{0}_{B}}
 (N^{(\nu)}_{R})^{X}_{\beta B}
 (N^{(u)*}_{L})^{u}_{BY}
 (N^{(u)}_{L})^{AY}_{u}
 (N^{(\nu)*}_{R})^{\alpha A}_{X}
,\\
\mathcal{B}^{(m,u\text{-}3)}_{\alpha\beta}
=&
 \frac{1}{4\sqrt{2} G_{F}}
 I_{4}
 \left(
 M_{\tilde{\chi}^{0}_{A}}, M_{\tilde{\chi}^{0}_{B}},
 m_{\tilde{\nu}_{X}}, m_{\tilde{u}_{Y}}
 \right)
 M_{\tilde{\chi}^{0}_{A}}
 M_{\tilde{\chi}^{0}_{B}}
 (N^{(\nu)}_{R})^{X}_{\beta B}
 (N^{(u)}_{R})^{Y}_{uB} 
 (N^{(\nu)*}_{R})^{\alpha A}_{X}
 (N^{(u)*}_{R})^{uA}_{Y}
,\\
\mathcal{B}^{(m,u\text{-}4)}_{\alpha\beta}
 =&
  -
 \frac{1}{8\sqrt{2} G_{F}}
  J_{4}
 \left(
 M_{\tilde{\chi}^{0}_{A}}, M_{\tilde{\chi}^{0}_{B}},
 m_{\tilde{\nu}_{X}}, m_{\tilde{u}_{Y}}
 \right)
 (N^{(\nu)}_{R})^{X}_{\beta B}
 (N^{(u)}_{R})^{Y}_{uB} 
 (N^{(\nu)*}_{R})^{\alpha A}_{X}
 (N^{(u)*}_{R})^{uA}_{Y}
,\\
\mathcal{B}^{(m,u\text{-}5)}_{\alpha\beta}
=&
 \frac{1}{8\sqrt{2} G_{F}}
  J_{4}
 \left(
 M_{\tilde{\chi}^{+}_{A}}, M_{\tilde{\chi}^{+}_{B}},
 m_{\tilde{l}_{X}}, m_{\tilde{d}_{Y}}
 \right)
 (C^{(\nu)}_{R})^{X}_{\beta B}
 (C^{(u)*}_{R})^{uB}_{Y}
 (C^{(u)}_{R})^{Y}_{uA}
 (C^{(\nu)*}_{R})^{\alpha A}_{X}
,\\
\mathcal{B}^{(m,u\text{-}6)}_{\alpha\beta}
=&
 -
 \frac{1}{4\sqrt{2} G_{F}}
I_{4}
 \left(
 M_{\tilde{\chi}^{+}_{A}}, M_{\tilde{\chi}^{+}_{B}},
 m_{\tilde{l}_{X}}, m_{\tilde{d}_{Y}}
 \right)
  M_{\tilde{\chi}^{+}_{A}}
 M_{\tilde{\chi}^{+}_{B}}
 (C^{(\nu)}_{R})^{X}_{\beta B}
 (C^{(u)*}_{L})^{uB}_{Y}
 (C^{(u)}_{L})^{Y}_{uA}
 (C^{(\nu)*}_{R})^{\alpha A}_{X}.
\end{align}

\subsection{For $\epsilon^{d}_{\mu\tau}$}
We consider the charged current interaction between neutrino beam and
the nucleon in the detector as a detection process.
It consists of the penguin contribution and the box contribution;
\begin{align}
\epsilon^{d}_{\mu\tau}
 =
 (\epsilon^{d}_{\mu\tau})_{\text{$W$-penguin}}
 +
 (\epsilon^{d}_{\mu\tau})_{\text{box}}.
\end{align}
The penguin contribution can be represented
as the complex conjugate of that for $\epsilon^{s}_{\tau \mu}$.
However, we must eliminate the diagrams which are already counted in 
the calculation of $\epsilon^{m}_{\mu\tau}$.
The detail is shown in Sec. \ref{sec:analytic-calc}. 
\begin{align}
(\epsilon^{d}_{\mu\tau})_{\text{$W$-penguin}}
=
\sum_{i = 3}^{7} \mathcal{A}^{(s\text{-}i)*}_{\tau\mu}.
\end{align}

The box contribution is calculated to be
\begin{align}
(\epsilon^{d}_{\mu\tau})_{\text{box}}
 =
 \sum_{i}
 \mathcal{B}^{(d\text{-}i)}_{\mu\tau},
\end{align}
where
\begin{align}
\mathcal{B}^{(d\text{-}1)}_{\alpha\beta}
=&
 \frac{1}{8\sqrt{2} G_{F}} 
 J_{4}
 \left(
 M_{\tilde{\chi}^{0}_{A}}, M_{\tilde{\chi}^{-}_{B}},
 m_{\tilde{\nu}_{X}}, m_{\tilde{u}_{Y}}
 \right)
 (C^{(l)}_{R})^{BX}_{\beta}
 (C^{(d)*}_{R})^{d}_{BY}
 (N_{R}^{(u)})^{Y}_{uA}
 (N^{(\nu)*})^{\alpha A}_{X}
,\\
\mathcal{B}^{(d\text{-}2)}_{\alpha\beta}
=& 
 \frac{1}{8\sqrt{2} G_{F}} 
 J_{4}
 \left(
 M_{\tilde{\chi}^{+}_{A}}, M_{\tilde{\chi}^{0}_{B}},
 m_{\tilde{l}_{X}}, m_{\tilde{d}_{Y}}
 \right)
 (N^{(l)}_{R})^{X}_{\beta B}
 (N^{(d) *}_{R})^{d B}_{Y}
 (C^{(u)}_{R})^{Y}_{uA}
 (C^{(\nu)*}_{R})^{\alpha A}_{X}
,\\
\mathcal{B}^{(d\text{-}3)}_{\alpha\beta}
 =&
 \frac{1}{4\sqrt{2} G_{F}}
 I_{4}
 \left(
 M_{\tilde{\chi}^{0}_{A}}, M_{\tilde{\chi}^{+}_{B}},
 m_{\tilde{\nu}_{X}}, m_{\tilde{d}_{Y}}
 \right)
 M_{\tilde{\chi}^{0}_{A}}
 M_{\tilde{\chi}^{+}_{B}}
 (C^{(l)}_{R})^{BX}_{\beta}
 (C^{(u)}_{R})^{Y}_{uB}
 (N^{(d)*}_{R})^{d A}_{Y}
 (N^{(\nu)}_{R})^{\alpha A}_{X}
,\\
\mathcal{B}^{(d\text{-}4)}_{\alpha\beta}
 =&
 \frac{1}{4\sqrt{2} G_{F}}
 I_{4}
 \left(
 M_{\tilde{\chi}^{-}_{A}}, M_{\tilde{\chi}^{0}_{B}},
 m_{\tilde{l}_{X}}, m_{\tilde{u}_{Y}}
 \right)
 M_{\tilde{\chi}^{-}_{A}} M_{\tilde{\chi}^{0}_{B}}
 (N^{(l)}_{R})^{X}_{\beta B}
 (N^{(u)}_{R})^{Y}_{uB}
 (C_{R}^{(d)*})^{d}_{AY}
 (C^{(\nu)*}_{R})^{\alpha A}_{X}
.
\end{align}

%
%
\newcommand{\Journal}[4]{{ #1} {\bf #2} {(#3)} {#4}}
\newcommand{\APJ}{Ap. J.}
\newcommand{\CJP}{Can. J. Phys.}
\newcommand{\EPJ}{Eur. Phys. J.}
\newcommand{\MPL}{Mod. Phys. Lett.}
\newcommand{\NC}{Nuovo Cimento}
\newcommand{\NP}{Nucl. Phys.}
\newcommand{\PL}{Phys. Lett.}
\newcommand{\PR}{Phys. Rev.}
\newcommand{\PRep}{Phys. Rep.}
\newcommand{\PRL}{Phys. Rev. Lett.}
\newcommand{\PTP}{Prog. Theor. Phys.}
\newcommand{\SJNP}{Sov. J. Nucl. Phys.}
\newcommand{\ZP}{Z. Phys.}
\newcommand{\EUR}{Eur. Phys. J.}


\end{document}